\documentclass[a4paper,10pt]{article}
\usepackage{jheppub}
\usepackage{amsfonts,amsmath,amssymb,graphicx,color}
\usepackage{float}
\usepackage{ulem}
\usepackage{subcaption}
\usepackage{tikz}
\usetikzlibrary{matrix,shapes,arrows,positioning,chains}

\DeclareMathOperator{\sgn}{sgn}
\def\gap#1{\vspace{#1 ex}}

\def\csch{{\rm cosech}}

\title{\textbf{Thermalization and non-monotonic entanglement growth in an exactly solvable model}}

\author[a]{Shruti Paranjape}\emailAdd{sparanjape@ucdavis.edu}

\author[b]{and Nilakash Sorokhaibam}\emailAdd{phy\_sns@tezu.ernet.in}

\affiliation[a]{Center for Quantum Mathematics and Physics (QMAP),\\
Department of Physics, University of California, Davis, CA 95616, USA}

\affiliation[b]{Department of Physics, Tezpur University, Tezpur, 784028, Assam, India}

\abstract{We study quantum quenches and subsequent non-equilibrium dynamics of free Dirac fermions in 1+1 spacetime dimensions using time dependent mass. The final state is a normalized boundary state which is called generalized Calabrese-Cardy (gCC) state and the system thermalizes to a generalized Gibb's Ensemble(GGE). We can also tune the initial states so that the final states are exact Calabrese-Cardy (CC) state and special gCC states. The system in the CC state thermalizes to a Gibb's ensemble. We derive closed-form analytic expressions for the growth of entanglement entropy of subsystems consisting of arbitrary number of disjoint intervals in CC state. We show that the entanglement entropy of a single interval grows monotonically before saturation. In case of certain gCC states, for particular charges, the entanglement entropy of a single interval grows non-monotonically when the effective chemical potential is increased beyond a critical value. We argue that the non-monotonic growth of entanglement entropy is due to increase in long range correlation and decrease in short range correlation at early times.}

\begin{document}

\maketitle

\section{\label{sec:intro}Introduction and Summary}

Thermalization in quantum systems has been a topic of great significance \cite{Polkovnikov:2010yn, gogolin2015equilibration}. Using AdS/CFT correspondence, it has also been linked to black hole formation \cite{Bhattacharyya:2009uu,Chesler:2008hg,Balasubramanian_2011,Liu:2013qca}. Closely related to thermalization (equilibration in general), the study of time evolution of a system after a quantum quench has also been a topic of intense research interest \cite{PhysRevA.69.053616, Calabrese:2006quench}. Quantum quench is the process in which the parameters of the Hamiltonian of a system are changed with time. The initial state is usually taken to be the ground state of the initial Hamiltonian.\footnote{In this paper, `initial state' means the initial state before the quantum quench while `final state' means the state immediately after the quantum quench process ended.} It is experimentally easier to prepare a gapped system in the ground state by sufficiently cooling the system. The final state is usually a non-equilibrium state of the final Hamiltonian. In the long time limit, if the system appears to be in a thermal state, in the sense that the expectation values of observables are same as the expectation values in a thermal ensemble, then we say that the system has thermalized. Note that the system is always in a pure state if we started from a pure state.

In this paper, we work in 1+1 dimensional spacetime. We will also restrict ourself to critical quantum quenches in which the final Hamiltonian is a critical Hamiltonian, i.e., the corresponding theory is a 1+1 dimensional conformal field theory (CFT). We will also consider infinite systems, i.e., the spatial extent of the system is the largest lengthscale in all our discussion. More specifically, we will be considering free Dirac fermions in which starting from a certain pure state in the massive theory, the mass is set to zero as a function of time. In more general theories, starting from the ground state of a gapped theory, it has been proposed \cite{Calabrese:2005in,Calabrese:2006rx} that the state obtained after the critical quench is a normalized boundary state called Calabrese-Cardy(CC) state which has the form
\begin{equation}
|\Psi_{CC}\rangle=e^{-\kappa_2 H}|B\rangle\
\label{CC_state_def}
\end{equation}
where $\kappa_2$ is a scale set by the initial gap and the other scales of the quench process, $H$ is the Hamiltonian of the CFT and $|B\rangle$ is a conformally invariant boundary state. It has been shown that such a state thermalizes to a thermal ensemble with temperature $T=1/\beta=1/(4\kappa_2)$. Incredibly, this result was shown using conformal symmetry without considering any specific CFT. This result has also been generalised to the case in which the final theory has other conserved charges of local currents \cite{Mandal:2015jla}. The corresponding ansatz for the state after quench from ground state is a generalized Calabrese-Cardy(gCC) state which has the form 
\begin{equation}
|\Psi_{gCC}\rangle=e^{-\kappa_2 H-\kappa_4 W_4-\kappa_6 W_6-...}|B\rangle\
\label{gCC_state_def}
\end{equation}
where again the parameters $\kappa_2, \kappa_4, \kappa_6,\cdots$ are given by the initial gap and other scales in the quench process, e.g. $\delta t=1/\rho$ the time taken to set the mass to zero, and $W_4, W_6, \cdots$ are the conserved charges of local currents \cite{Bakas:1990ry,Pope:1991ig}. In this case also, it has been shown that the state thermalizes into a generalized Gibb's Ensemble(GGE) with the density matrix $e^{-\beta H -\mu_4 W_4 - \mu_6 W_6-...}$ where the corresponding temperature and chemical potentials are $T=1/\beta=1/(4\kappa_2)$, $\mu_4=4\kappa_4$, $\mu_6=4\kappa_6, \cdots$.

The gCC state ansatz has been shown to be true for mass quenches in free scalar and free fermionic theories \cite{Mandal:2015kxi}. The mass is the time-dependent parameter. The mass is turned off in the far future so that the final theory is a CFT. Starting from the ground state of the massive theories, the quenched states obtained are of the gCC form with infinite number of charges $W_{2n}$ with $n\in\mathbb{N}$ ($W_2=H$). $W_{2n}=\sum |k|^{2n-1}d^\dagger_kd_k$, where $d^\dagger_k$ and $d_k$ are the annihilation and creation operators, are the familiar bilinear $W$ charges of free scalar and fermionic theories \cite{Bakas:1990ry,Pope:1991ig}.\footnote{The normalization of the charges differ from the normalization in \cite{Bakas:1990ry,Pope:1991ig}.} It was also shown that starting from specially prepared squeezed states of the massive scalar theory, CC state and gCC state with finite number of charges can also be created. By calculating correlators, thermalization of these states were explicitly shown without using perturbative expansion and resummation.

\paragraph{Thermalization in free theories:} Thermalization of CC and gCC states (even in free theories, and integrable theories in general) is rather mysterious and seems a bit different from the conventional thermalization that one observes in highly chaotic systems \cite{Eberlein:2017wah,Bhattacharya:2018fkq,Samui:2020jli}. This is the reason why this phenomenon has been called `subsystem thermalization' in \cite{Mandal:2015kxi}. Thermalization in these theories to GGE in the long time limit is because the final states have only short range correlation. This follows from central limit theorem and the local nature of the initial and final Hamiltonians \cite{Rigol_2008nature}. Non-equilibrium dynamics of chaotic systems and integrable systems develop a stark contrast when we consider critical-to-critical quench \cite{Mandal:2015kxi} or bump quenches \cite{Bhattacharya:2018fkq}. These are finite perturbation for a finite duration of time, the initial and the final Hamiltonians are same. Chaotic systems still thermalize while integrable systems do not thermalize. In summary, CC states and gCC states in integrable systems thermalize because these states are special fine-tuned states. Even states in chaotic systems need certain constraint to thermalize - the states should be typical states \cite{Srednicki_1999}. But this is a very mild constraint on the variance of the energy compared to the fine-tuning of CC and gCC states in integrable systems.

Another important aspect which have got significant attention is diffusive dynamics and the associated hydrodynamic long-time tails after quantum quenches \cite{Lux_2014,Buchhold_2016,Rakovszky_2019,Alba_2021}. Diffusive dynamics could occur due to spatial inhomogeneity in excitations, say in our case, the mass is also space dependent. But for the critical quenches in 1+1 dimensions that we will be concentrating on there is no diffusive dynamics because the final theory is a 1+1 dimensional CFT \cite{Sotiriadis2008}. The hydrodynamic long-time tails could also arise even in spatially homogeneous excitations, the fluctuations could die down to equilibrium fluctuation pattern very slowly as a power law \cite{Lux_2014}. But again, such long-time tails do not exist in a free theory because there is no-mixing of the momentum modes.

Entanglement plays an important role in thermalization of pure states or thermal nature of pure states. Starting from a pure state of the initial Hamiltonian, the full system is always in a pure state so there is no thermal entropy. The entropy of a large subsystem\footnote{Subsystem size is much larger than lengthscale given by the effective inverse temperature.} after thermalization is entirely given by the entanglement entropy of the subsystem with its conjugate subsystem. In case of non-equilibrium dynamics, the entanglement entropy of a subsystem grows and saturates at the value given by the thermal state. The growth of entanglement entropy of a subsystem has been calculated (mostly numerically) in many systems, see e.g. \cite{Calabrese:2005in,Calabrese:2007-local,Sotiriadis_2008,PhysRevA.78.010306,PhysRevX.3.031015,Nezhadhaghighi:2014pwa,Rajabpour_2015, Coser_2014,Nahum:2016muy,Cotler_2016,Fr_rot_2018,A_2018a,A_2018b,Najafi_2018,Bertini_2018,De_Luca_2020,Bastianello_2020}. It has also been extensively examined in holographic systems \cite{AbajoArrastia:2010yt,Hartman:2013qma,Caputa:2013eka,Liu:2013qca,Kundu:2016cgh}.

Using bosonization, we derive analytic expression for entanglement entropy for subsystems consisting of arbitrary number of disjoint intervals in CC state and certain gCC states. For CC state, we can even write down a closed form expression. In CC state, we found that the entanglement entropy of a single interval grows monotonically without any assumption of high effective temperature.\footnote{It should be noted that the entanglement growth of subsystem consisting of multiple disjoint intervals does not grow monotonically.} At early time, the entanglement growth is quadratic in time, as expected in \cite{Liu:2013qca,Nezhadhaghighi:2014pwa,Kundu:2016cgh}. The long time limit is given by the well-known expression from CFT in a thermal ensemble, $S_A=\frac{c}{3}\log[\sinh(\pi r/4\kappa_2)]$, where for Dirac fermions $c=1$ and the effective inverse temperature $\beta =4\kappa_2$. It has earlier been shown that the entanglement entropy of a single interval grows linearly in the high effective temperature limit (or large subsystem size) before saturating abruptly to the final thermal value \cite{Calabrese:2005in}. Our expression for entanglement entropy agrees with this result. There has been attempts to explain black hole entropy as entanglement entropy \cite{Sorkin:1986mg,Sorkin:1997ja,Emparan:2006ni,Azeyanagi:2007bj,Solodukhin:2011gn}. Moreover, it is well known that black hole entropy obeys a second law which states that the area (or its generalization and correspondingly the black hole entropy) always increases \cite{Wall:2015raa,Bhattacharya:2019qal}. It appears to be more than a coincidence that the entanglement entropy of CC state grows monotonically, because according to AdS/CFT correspondence, a black hole corresponds to a thermal state in the boundary CFT. This is irrespective of the fact that we are considering a free theory while holographic CFTs are strongly interacting systems. The reason is that the non-equilibrium dynamics (not only long time limits) of entanglement entropy of a single interval in CC state of all CFTs have a universal behaviour because they are fixed by conformal symmetry. Only at very early times and at the time of saturation, the details of the theory is important \cite{Calabrese:2006quench}. On the other hand, the entanglement growth of subsystem consisting of multiple intervals in not universal. For example, it is different for integrable systems and for holographic systems. This can be starkly identified from the non-equilibrium dynamics of mutual information of two disjoint intervals in the two different theories. Despite this, one should note that the long time limit in all theories are given by the thermal values.

\begin{figure}
  \centering
  \begin{subfigure}{.5\textwidth}
  \includegraphics[width=.9\linewidth]{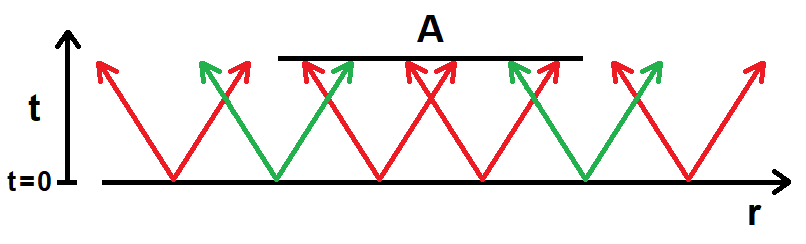}
  \caption{\small Low entanglement at early times.}
  \label{fig:low_early}
\end{subfigure}%
\begin{subfigure}{.5\textwidth}
  \centering
  \includegraphics[width=.9\linewidth]{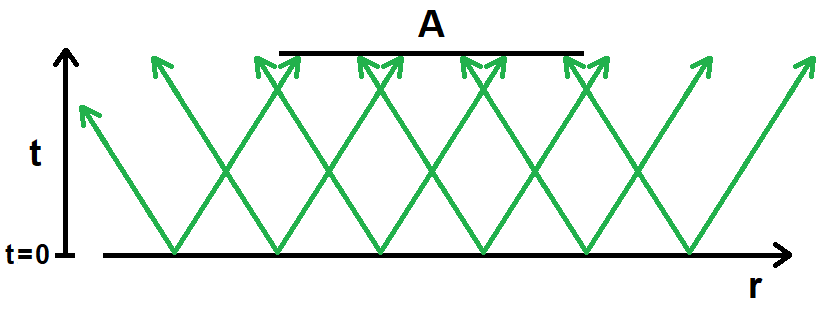}
  \caption{\small Maximal entanglement at saturation time.}
  \label{fig:maximal_saturation}
\end{subfigure}
\begin{subfigure}{.5\textwidth}
  \centering
  \includegraphics[width=.9\linewidth]{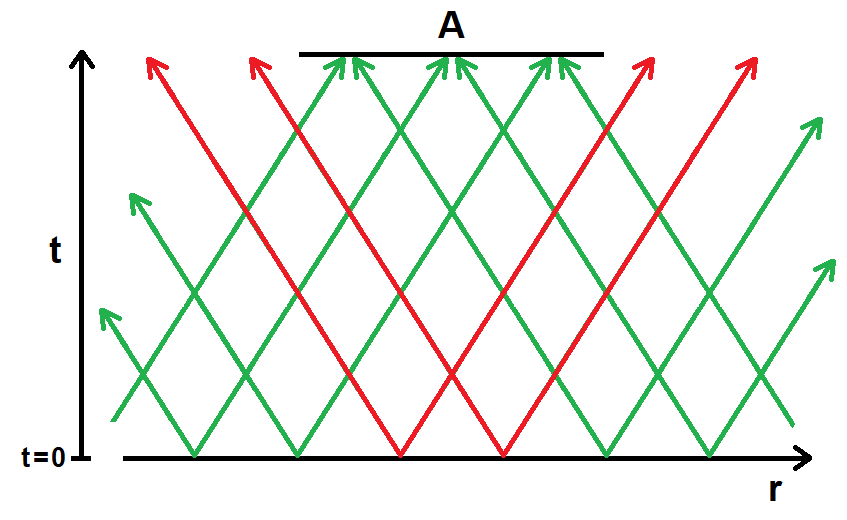}
  \caption{\small Maximal entanglement at late time.}
  \label{fig:maximal_late}
\end{subfigure}
\caption{\small Quasiparticle picture of entanglement propagation. Quasiparticle pairs in green entangle subsystem $A$ with the rest of the system while the red ones do not contribute to entanglement.}
\label{fig:quasi_pic}
\end{figure}

The growth of entanglement entropy in integrable systems has been phenomenologically explained using a quasiparticle picture \cite{Calabrese:2005in,Calabrese:2007-local,Sotiriadis_2008,PhysRevA.78.010306,PhysRevX.3.031015,Nezhadhaghighi:2014pwa, Rajabpour_2015, Coser_2014, Cotler_2016, Fr_rot_2018, A_2018a, A_2018b, Najafi_2018,  Bertini_2018, De_Luca_2020, Bastianello_2020, Calabrese_2020, kudlerflam2020quasiparticle}. As shown in figure \ref{fig:low_early}, it is assumed that coherent pairs of quasiparticles are emitted with opposite momenta immediately after the quench process (or at time $t=0$ for the CC state or gCC states). It is assumed that the quasiparticles do not scatter and do not decay. So as soon as one particle reaches a subsystem A and its entangled partner reaches subsystem B, A and B are entangled. In the simplest case, for conformal field theories at high effective temperature (or large subsystem size), the quasiparticles propagate along the light-cone with a speed $v_p=1$ \cite{Calabrese_2020}. The post-quench state is also always in the form of a squeezed state which gives a mathematical representation of this picture. This leads to linear growth of entanglement entropy. The saturation (thermalization) of the entanglement entropy is also explained by this picture as when the subsystem has accommodated the maximum number of entangled particles as shown in figure \ref{fig:maximal_saturation}. For a single interval of length $r$, this maximum is reached at time $t=r/(2 v_p)$. Further time evolution represented in Figure \ref{fig:maximal_late} do not lead to increase or decrease in the entanglement entropy. The quasiparticle picture has also been modified to accommodate interacting, and chaotic theories \cite{Bastianello_2020,kudlerflam2020quasiparticle}.

The monotonic growth of entanglement entropy for a single interval is not always the case if we turn on other conserved charges. In case of gCC states with certain charges, we found that the entanglement entropy of a single interval can grow non-monotonically. The entanglement entropy increases rapidly before decreasing to some extent and settles down to its final value. It should be noted that we have to zoom in near the saturation time to see this behaviour. These states are gCC states with non-linear $W$ charges. We also show that the non-monotonic growth of entanglement entropy happens when the effective chemical potential is increased beyond a critical value. This is akin to a dynamical phase transition \cite{Heyl_2018}, here the rate of change of entanglement entropy with time becomes zero at finite time when the effective chemical potential is greater than the critical value. We argue that the non-monotonic growth is due to increase in long range correlation and decrease in short range correlation at early times. But, the quasiparticle picture fails to reproduce this result even after modifications taking into account the changes in early time quantum correlation. So, we believe that the non-monotonic growth is due to intricate interplay of the increase in long range correlation and decrease in short range correlation. This also suggests that the more familiar bilinear $W_{2n}$ charges of free fermionic theory could also give rise to non-monotonic growth of entanglement entropy because of the higher derivatives involved which would give rise to longer range correlation at early times. Non-monotonic entanglement growth consisting of an initial dip around the quench time (unlike in our present case) has also been observed in a holographic set-up in \cite{Bai:2014tla}.

\gap1
\noindent In summary, the main results of the present work are:
\begin{enumerate}
\item For CC state, we are able to write down a closed-form expression of the growth of entanglement entropy of subsystems consisting of arbitrary number of disjoint intervals. The entanglement growth for a single interval is monotonically increasing. The early time growth is quadratic in time. At high effective temperature, the entanglement growth is linear in time. The stationary limit is, as expected, the entanglement entropy of a single interval in thermal ensemble.

\item We found that the growth of entanglement entropy of a single interval in certain gCC state with non-linear $W_{2n}$ charge can become non-monotonic. The maximum value is not the long-time thermal entropy for the subsystem. The non-monotonic growth of entanglement entropy happens when the effective chemical potential of the non-linear $W_{2n}$ charge is beyond a critical value.

\item We provide evidences supporting our claim that the non-monotonic growth of entanglement entropy is due to increase in long range correlation and decrease in short range correlation at early times.
\end{enumerate}

\gap1

\noindent The outline of the paper is as follows:

\gap1

\noindent In section \ref{theo}, we solve the Dirac equation with time-dependent mass and from explicit solutions for a specific mass profile, we calculate the Bogoliubov coefficients for the transformation between the massive and massless modes. In section \ref{qstates}, we find the final state after the quench starting from the ground state and a few squeezed states of our interest. In sections \ref{ed} and \ref{crf}, we calculate energy density and some correlators in the different quenched states that we obtained. The entanglement growth of a single interval and multi-partite subsystems in CC state is explicitly calculated in section \ref{secEE_CC}. In section \ref{secEE_gCC}, we show that the entanglement growth of a single interval in a particular gCC state is non-monotonic when the effective chemical potential is beyond a critical value. In section \ref{secEE_gCC_arg}, we argue that the non-monotonic growth of entanglement entropy is due to long range correlation at early times. Section \ref{cond} is conclusions. The appendix contains details that we have omitted in the main sections.

\section{\label{theo}Free Dirac fermions with time-dependent mass}
The action for Dirac fermions with time-dependent mass is
\begin{eqnarray}
S=-\int dx^2\left[i\bar{\Psi}\gamma^\mu\partial_\mu\Psi-m(t)\bar{\Psi}\Psi\right]\
\label{action}
\end{eqnarray}
The equation of motion (EOM) is
\begin{eqnarray}
 \left[ i\gamma^0\partial_t-i\gamma^1\partial_x-m(t)\right]\Psi(x,t)=0\
 \label{eom}
\end{eqnarray}
and we are interested in the solvable mass profile\cite{Duncan:1977fc, Das:2014hqa}
\begin{eqnarray}
m(t)=m[1-\tanh(\rho t)]/2\
 \label{mass}
\end{eqnarray}
$m$ is the initial mass and $\rho$ is the only scale of the quench process. $\rho\to\infty$ is the sudden limit in which the mass is set to zero suddenly - much faster than any other length scale in the theory.
It is easier to solve (\ref{eom}) in the Dirac basis in which $\gamma_0$ is diagonal. Since the system is translation invariant in the  spatial $x$-direction, the solution ansatz is
\begin{eqnarray}
 \Psi(x,t)=\left[\gamma^0\partial_t-\gamma^1\partial_x-im(t)\right]e^{\pm i kx}\Phi(t)\
 \label{spinor_antz}
\end{eqnarray}
Substitution in the EOM gives,
\begin{eqnarray}
 \left[\partial^2_t+k^2+m(t)^2-i\gamma^0\dot{m}(t)\right]e^{\pm i kx}\Phi(t)=0\nonumber\
\end{eqnarray}
where $\dot{m}(t)=\partial_tm(t)$.

$\Phi(t)$ is solved in the eigenbasis of $\gamma^0$. For the two eigenvalues of $\gamma^0$ (1 and -1), the two solutions $\phi_+(t)$ and $\phi_-(t)$ are given by,
\begin{eqnarray}
 &&\left[\partial^2_t+k^2+m(t)^2-i\dot{m}(t)\right]\phi_+(t)=0\nonumber\\
 &&\left[\partial^2_t+k^2+m(t)^2+i\dot{m}(t)\right]\phi_-(t)=0\
 \label{pppm}
\end{eqnarray}
where $\Phi(t)=\begin{bmatrix} \phi_+(t) & \phi_-(t)  \end{bmatrix}^\text{T}$. The eigenstates of $\gamma^0$ are $u_0=\begin{bmatrix} 1 \\ 0  \end{bmatrix}$ and $v_0=\begin{bmatrix} 0 \\ 1 \end{bmatrix}$, they are the spinors in the rest frame.

For the mass profile (\ref{mass}), there are two important bases of solutions in which we are interested in. The first one is the `in' basis in which the two independent solutions of the second order linear differential equations become different single frequency modes in the $t\to -\infty$ limit. In other words, one solution becomes the negative energy mode and the other solution becomes the positive energy mode. Similarly, there is also an `out' basis of solutions in which one solution becomes the negative energy mode and the other becomes the positive energy mode in the $t\to\infty$ limit. Accordingly, we will also have different `in' and `out' creation and annihilation operators.
Consider the solutions of (\ref{pppm}) in the two bases to be
\begin{eqnarray}
&&\phi_{\pm}(t,k)=\phi_{in,\pm p}(t,k)+\phi_{in,\pm m}(t,k)\\
&&\phi_{\pm}(t,k)=\phi_{out,\pm p}(t,k)+\phi_{out,\pm m}(t,k)\
\end{eqnarray}
where the limits are
\begin{eqnarray}
&&\lim_{t\to-\infty} \phi_{in,\pm p}(t,k)=e^{-i\omega_{in} t}, \quad \lim_{t\to-\infty}\phi_{in,\pm m}(t,k)=e^{i\omega_{in} t}\nonumber\\
&&\lim_{t\to\infty} \phi_{out,\pm p}(t,k)=e^{-i\omega_{out} t}, \quad \lim_{t\to\infty}\phi_{out,\pm m}(t,k)=e^{i\omega_{out} t}\nonumber\
\end{eqnarray}
where `p' means {\it positive energy} and `m' means {\it negative energy}. The above solutions are the same but written in two different bases for simplicity in the appropriate time limits, they are related by Bogoliubov transformations.\\
But from (\ref{pppm}), we see that the equations of $\phi_+$ and $\phi_-$ are the complex conjugates of each other, so
\begin{eqnarray}
 &&\phi_{in,+p}(t,k)=\phi^*_{in,-m}(t,k), \quad \phi_{in,+m}(t,k)=\phi^*_{in,-p}(t,k)\\
 \label{conj1}
 &&\phi_{out,+p}(t,k)=\phi^*_{out,-m}(t,k), \quad \phi_{out,+m}(t,k)=\phi^*_{out,-p}(t,k)\
 \label{conj2}
\end{eqnarray}
The Bogoliubov transformations are
\begin{eqnarray}
 \phi_{in,+p}(t,k)&=&\alpha'_+(k)\phi_{out,+p}(t,k)+\beta'_+(k)\phi_{out,+m}(t,k)\nonumber\\
 \label{bop}&=&\alpha'_+(k)\phi_{out,+p}(t,k)+\beta'_+(k)\phi^*_{out,-p}(t,k)\\
 \phi_{in,-p}(t,k)&=&\alpha'_-(k)\phi_{out,-p}(t,k)+\beta'_-(k)\phi_{out,-m}(t,k)\nonumber\\
 &=&\alpha'_-(k)\phi_{out,-p}(t,k)+\beta'_-(k)\phi^*_{out,+p}(t,k)\ \label{bom}
\end{eqnarray}
where $\alpha'_{\pm}(k)$ an $\beta'_{\pm}(k)$ are actually functions of $|k|$, since the equations of motion have only $k^2$ terms.

Now, suppressing the basis labels `in' and `out' since they apply to both bases, we write the $u_0$ part of $\Psi(x,t)$ as (upto normalization)
\begin{eqnarray}
 \tilde{U}(x,t;k)=\left[ \gamma^0\partial_t+\gamma^1\partial_x -im(t)\right]e^{ikx}\phi_{+p}(t)\begin{bmatrix} 1 \\ 0  \end{bmatrix}\
 \label{Uxt}
\end{eqnarray}
And the $v_0$ part of $\Psi(x,t)$ as
\begin{eqnarray}
 \tilde{V}(x,t;k)&=&\left[ \gamma^0\partial_t+\gamma^1\partial_x -im(t)\right]e^{-ikx}\phi_{-m}(t)\begin{bmatrix} 0 \\ 1  \end{bmatrix}\nonumber\\
 &=&\left[ \gamma^0\partial_t+\gamma^1\partial_x -im(t)\right]e^{-ikx}\phi^*_{+p}(t)\begin{bmatrix} 0 \\ 1  \end{bmatrix}\
 \label{Vxt}
\end{eqnarray}
We can define the spinors as (upto normalization)
\begin{eqnarray}
 \tilde{u}(t,k)&=&\frac{1}{e^{ikx}\phi_{+p}(t)}\tilde{U}(x,t;k)\nonumber\\
 \tilde{v}(t,k)&=&\frac{1}{e^{-ikx}\phi_{-m}(t)}\tilde{V}(x,t;k)\nonumber\
\end{eqnarray}
With proper normalization, the final Dirac fermion mode expansion is
\begin{eqnarray}
 \Psi(x,t)&=&\int \frac{dk}{2\pi}\frac{1}{\sqrt{2\omega}}\left[a_k U(x,t;k)+b^\dagger_k V(x,t;k)\right]\nonumber\\
 &=&\int \frac{dk}{2\pi}\frac{1}{\sqrt{2\omega}}\left[a_k u(t;k)e^{ikx}\phi_{+p}(t)+b^\dagger_k v(t;k)e^{-ikx}\phi_{-m}(t)\right]\
 \label{modes}
\end{eqnarray}

\subsection{Bogoliubov transformation of oscillators}
The initial mass is taken to be $\lim_{t\to -\infty}m(t)=m$. It is convenient to take the final mass $\lim_{t\to\infty}m(t)$ to be some $m_{out}$, because of the spinor convention (in P\&S), although we are interested in $m_{out}=0$.

With time-dependent mass, as mentioned above, the spinors are functionals of $m(t)$, but their normalizations are constants or else they will not solve the Dirac equations. So, we have to differentiate between `in' spinors and `out' spinors. Taking this into account, the mode expansion of $\Psi(x,t)$ starting from `in' basis to `out' basis is
\begin{eqnarray}
 \Psi(x,t)&=&\int \frac{dk}{2\pi}\frac{1}{\sqrt{2\omega_{in}}}\left[a_{in,k} u_{in}(k,m) \phi_{in,+p}(t,k) e^{ikx}+b^\dagger_{in,k} v_{in}(k,m)\phi^*_{in,+p}(t,k)e^{-ikx}\right]\nonumber\\
 &=&\int \frac{dk}{2\pi}\frac{1}{\sqrt{2\omega_{in}}}\Huge{[} \{\alpha'_+(k)a_{in,k} u_{in}(k,m)\phi_{out,+p}(t,k)+b^\dagger_{in,-k} v_{in}(-k,m)\beta'^*_+(k)\phi_{out,-p}(t,k)\} e^{ikx}\nonumber\\
 &&\quad+\{\alpha'^*_+(k)b^\dagger_{in,k} v_{in}(k,m)\phi^*_{out,+p}(t,k)+a_{in,-k} u_{in}(-k,m)\beta'_+(k)\phi^*_{out,-p}(t,k)\}e^{-ikx}\huge{]}\nonumber\\
\end{eqnarray}
where we have used the facts that the $k$ integral is from $-\infty$ to $\infty$ and $\alpha'_{\pm}$, $\beta'_{\pm}$ and $\phi_{\pm p}$ are functions of $|k|$. In $t\to\infty$ limit, $m(t)\to m_{out}$, so,
\begin{eqnarray}
 \lim_{t\to\infty}\Psi(x,t)&=&\int \frac{dk}{2\pi}\frac{1}{\sqrt{2\omega_{out}}}\sqrt{\frac{\omega_{out}}{\omega_{in}}}\huge{[} \{\alpha'_+(k)a_{in,k} u_{in}(k,m_{out})+b^\dagger_{in,-k} v_{in}(-k,m_{out})\beta'^*_+(k)\} e^{-ik\cdot x}\nonumber\\
 &&\quad+\{\alpha'^*_+(k)b^\dagger_{in,k} v_{in}(k,m_{out})+a_{in,-k} u_{in}(-k,m_{out})\beta'_+(k)\}e^{ik\cdot x}\huge{]}\nonumber\
\end{eqnarray}
Comparing with the mode expansion in the `out' solution basis in the same limit $t\to\infty$,
\begin{eqnarray}
 \lim_{t\to\infty}\Psi(x,t)&=&\int \frac{dk}{2\pi}\frac{1}{\sqrt{2\omega_{out}}}\left[ a_{out,k} u_{out}(k,m_{out}) \phi_{out,+p}(t,k) e^{ikx}+b^\dagger_{out,k} v_{out}(k,m_{out})\phi^*_{out,+p}(t,k)e^{-ikx}\right]\nonumber\
\end{eqnarray}
we get the Bogoliubov transformations of the creation and annihilation operators.
\begin{eqnarray}
\label{aout} a_{out,k}&=&\alpha_+(k)a_{in,k}+b^\dagger_{in,-k}\, \chi(k,m_{out})\beta^*_+(k)\\
 b^\dagger_{out,k}&=&\alpha^*_+(k)b^\dagger_{in,k} +a_{in,-k}\, \tilde{\chi}(k,m_{out})\beta_+(k)\
 \label{bdout}
\end{eqnarray}
where $\alpha_+(k)=\sqrt{\frac{\omega_{out}(\omega_{out}+m_{out})}{\omega_{in}(\omega_{in}+m_{in})}}\,\alpha'_+$ and $\beta_+(k)=\sqrt{\frac{\omega_{out}(\omega_{out}+m_{out})}{\omega_{in}(\omega_{in}+m_{in})}}\,\beta'_+(k)$. Using (\ref{norspinor})
\begin{eqnarray}
 \chi(k,m_{out}) &=& \frac{1}{2m_{out}}\sqrt{\frac{\omega_{in}+m_{in}}{\omega_{out}+m_{out}}}\,\bar{u}_{out}(k,m_{out},\omega_{out})v_{in}(-k,m_{out},-\omega_{out})\nonumber\\
 &=&\text{sgn}(k)\qquad\qquad \text{when}\quad m_{out}\to0\
 \label{chi}
\end{eqnarray}
where we have to be careful that $v_{in}(k,m_{out})$ is a functional of the accompanying mode, which is $\sim e^{-i\omega_{out}t}$ in the above case. Similarly, with $m_{out}\to0$,
\begin{eqnarray}
\tilde{\chi}(k)&=&-\frac{1}{2m_{out}}\sqrt{\frac{\omega_{in}+m_{in}}{\omega_{out}+m_{out}}}\,\bar{v}(k,m_{out},\omega_{out})u(-k,m_{out},\omega_{out})=\text{sgn}(k) \
\label{chit}
\end{eqnarray}
taking into account the normalization of $\bar{v}_{out}v_{out}=-2m_{out}$. Inverting (\ref{aout}) and (\ref{bdout}), we get
\begin{eqnarray}
\label{ain} a_{in,k}&=&\alpha^*_+(k)a_{out,k}-\text{sgn}(k)\beta^*_+(k)b^\dagger_{out,-k}\\
b^\dagger_{in,-k}&=&\alpha_+(k)b^\dagger_{out,-k}+\text{sgn}(k)\beta_+(k)a_{out,k}\
\label{bdin}
\end{eqnarray}
From here on, we will suppress the subscript `out' on creation and annihilation operators, so $a_{out,k}=a_k$, similarly for $b_{out,k}$ and their Hermitian conjugates. Also, since $\chi(k)$ and $\tilde{\chi}(k)$ are simple sign functions, with a slight abuse of the nomenclature, we will call $\alpha_+(k)$ and $\beta_+(k)$ as the Bogoluibov coefficients.
Moreover, $\chi(k)^2$ and $\tilde{\chi}(k)^2$ are identically equal to 1. So, the fermionic anti-commutation relations of the `in' and `out' operators constraint the Bogoluibov coefficients as
\begin{eqnarray}
\label{ab1}|\alpha_+(k)|^2+|\beta_+(k)|^2=1\ 
\end{eqnarray}

\subsection{Explicit solutions}
In the `in' basis, for our choice of mass profile, the solutions are
\begin{eqnarray}
\phi_{in,+p} &=& e^{-i t \left(\omega+m\right)} \left(e^{-2 \rho  t}+1\right)^{-\frac{i m}{2 \rho }} \ _2F_1\left(\frac{i\left(|k|-m-\omega\right)}{2 \rho },-\frac{i \left(|k|+m+\omega\right)}{2 \rho };1-\frac{i \omega}{\rho };-e^{2 t\rho }\right)\nonumber\\
\phi_{in,-m} &=& e^{i t \left(\omega-m\right)} \left(e^{-2 \rho  t}+1\right)^{-\frac{i m}{2 \rho }}\ _2F_1\left(\frac{i\left(-|k|-m+\omega\right)}{2 \rho },\frac{i \left(|k|-m+\omega\right)}{2 \rho };\frac{i \omega}{\rho }+1; -e^{2 t\rho}\right)\nonumber\
\end{eqnarray}
where $\omega = \sqrt{k^2+m^2}$. While in the `out' basis, the solutions are
\begin{eqnarray}
\phi_{out,+p} &=& e^{-i |k| t} \left(e^{-2\rho t}+1\right)^{-\frac{i m}{2 \rho }} \ _2F_1\left(\frac{i |k|-i m+i \omega}{2 \rho },\frac{i |k|-i m-i \omega}{2 \rho };1+\frac{i |k|}{\rho };-e^{-2\rho t}\right)\nonumber\\
\phi_{out,-m} &=& e^{i |k| t} \left(e^{-2\rho t}+1\right)^{-\frac{i m}{2 \rho }} \ _2F_1\left(\frac{-i |k|-i m+i \omega}{2 \rho },\frac{-i|k|-i m-i \omega}{2 \rho};1-\frac{i |k|}{\rho };-e^{-2\rho t}\right)\nonumber\
\end{eqnarray}

Using the properties of confluent hypergeometric functions $_2F_1$ given in \cite{Abramowitz}, the Bogoliubov coefficients of the frequency modes as defined in (\ref{bop}) are
\begin{eqnarray}
\label{alphap}
\alpha'_+ &=& \frac{\Gamma \left(-\frac{i |k|}{\rho }\right) \Gamma \left(1-\frac{i \omega}{\rho }\right)}{\Gamma \left(-\frac{i\left(|k|+m+\omega\right)}{2 \rho }\right) \Gamma \left(1+\frac{-i |k|+i m -i \omega}{2 \rho }\right)}\\
\label{betap}
\beta'_+ &=& \frac{\Gamma \left(\frac{i |k|}{\rho }\right) \Gamma \left(1-\frac{i \omega}{\rho }\right)}{\Gamma \left(1-\frac{i\left(-|k|-m+\omega\right)}{2 \rho }\right) \Gamma \left(-\frac{i \left(-|k|+m+\omega\right)}{2 \rho }\right)}
\end{eqnarray}
In the sudden limit($\rho\to\infty$). the Bogoliubov coefficients of the frequency modes are
\begin{eqnarray}
\alpha'_+(k) &=& \frac{|k|+m_{in}+\sqrt{k^2+m^2}}{2|k|}\\
\beta'_+(k) &=& \frac{|k|-m_{in}-\sqrt{k^2+m^2}}{2|k|}\
\end{eqnarray}
As mentioned above, for a quench starting from the ground state of the massive theory, the naive sudden limit gives a non-normalizable state in the massless theory \cite{Mandal:2015kxi}. The problem arises only for a quench starting from the ground state. In case the quench is starting from squeezed states of our interest, the naive sudden limit given above works well.
As defined in (\ref{aout}) and (\ref{bdout}), the Bogoluibov coefficients of the oscillator modes differ from $\alpha'_+(k)$ and $\beta'_+(k)$ by an overall factor.
\begin{eqnarray}
\label{alpha}
\alpha_+ &=& \sqrt{1-\frac{m}{\sqrt{k^2+m^2}}}\frac{\Gamma \left(-\frac{i |k|}{\rho }\right) \Gamma \left(1-\frac{i \omega}{\rho }\right)}{\Gamma \left(-\frac{i\left(|k|+m+\omega\right)}{2 \rho }\right) \Gamma \left(1+\frac{-i |k|+i m -i \omega}{2 \rho }\right)}\\
\label{beta}
\beta_+ &=& \sqrt{1-\frac{m}{\sqrt{k^2+m^2}}}\frac{\Gamma \left(\frac{i |k|}{\rho }\right) \Gamma \left(1-\frac{i \omega}{\rho }\right)}{\Gamma \left(1-\frac{i\left(-|k|-m+\omega\right)}{2 \rho }\right) \Gamma \left(-\frac{i \left(-|k|+m+\omega\right)}{2 \rho }\right)}
\end{eqnarray}
In the sudden limit, they are
\begin{eqnarray}
\label{alphas} \alpha_+(k) &=& \sqrt{1-\frac{m}{\sqrt{k^2+m^2}}}\, \frac{|k|+m+\sqrt{k^2+m^2}}{2|k|}\\
\label{betas} \beta_+(k) &=& \sqrt{1-\frac{m}{\sqrt{k^2+m^2}}}\, \frac{|k|-m-\sqrt{k^2+m^2}}{2|k|}\
\end{eqnarray}

For completeness, the expressions of $\alpha'_-$ and $\beta'_-$ in (\ref{bom}) for our particular quench protocol are
\begin{eqnarray}
\alpha'_- &=& \frac{\Gamma \left(-\frac{i \left| k\right| }{\rho }\right) \Gamma\left(1-\frac{i \omega}{\rho }\right)}{\Gamma \left(-\frac{i\left(\left| k\right| -m+\omega\right)}{2 \rho }\right) \Gamma \left(1-\frac{i \left(\left| k\right| +m+\omega \right)}{2 \rho }\right)}\\
\beta'_-&=& \frac{\Gamma \left(\frac{i \left| k\right| }{\rho }\right) \Gamma \left(1-\frac{i \omega}{\rho }\right)}{\Gamma \left(\frac{i\left(\left| k\right| +m-\omega\right)}{2 \rho }\right) \Gamma \left(1-\frac{i \left(-\left| k\right|+m+\omega\right)}{2 \rho }\right)}
\end{eqnarray}

\section{\label{qstates}Quenched states}

\subsection{From ground state}
Starting from the ground state of the massive theory $|\Psi\rangle=|0,in\rangle$, using Eq (\ref{ain}), the state in terms of `out' operators is given by
\begin{align}
a_{in,k}|\Psi\rangle=0 \quad&\Rightarrow \left[\alpha^*_+(k)a_{k}-\text{sgn}(k)\beta^*_+(k)b^\dagger_{-k}\right]|\Psi\rangle=0\nonumber\\
\label{psig}&\Rightarrow |\Psi\rangle = e^{\sum_k \text{sgn(k)}\gamma(k) a^\dagger_{k}b^\dagger_{-k}}|0\rangle\\
\text{where} &\quad\gamma(k)=\frac{\alpha^*_+(k)}{\beta^*_+(k)}\
\label{gg}
\end{align}
where we have taken $|0\rangle$ to be the ground state of `out' oscillators. Using the Baker-Campbell-Hausdorff(BCH) formula derived in appendix (\ref{bch}), the above state can be written in gCC form. For the particular mass profile (\ref{mass}), $\alpha_+(k)$ and $\beta_+(k)$ are given in (\ref{alpha}) and (\ref{beta}). The gCC form which was first obtained in \cite{Mandal:2015kxi} is
\begin{eqnarray}
\label{psigg}|\Psi\rangle =e^{-\kappa_2 H-\kappa_4W_4-\kappa_6W_6-...}|D\rangle\
\end{eqnarray}
where
\begin{gather}
\kappa_2=\frac{1}{2 m}+\frac{\pi ^2 m}{12 \rho ^2}+\frac{1}{m}\mathcal{O}\left(\frac{m}{\rho}\right)^3, \quad \kappa_4=-\frac{1}{12 m^3}+\frac{\pi ^2}{24 m \rho ^2}+\frac{1}{m^3}\mathcal{O}\left(\frac{m}{\rho}\right)^3,\nonumber\\
\kappa_6=\frac{3}{80 m^5}-\frac{\pi ^2}{96 m^3 \rho ^2}+\frac{1}{m^5}\mathcal{O}\left(\frac{m}{\rho}\right)^3,...\
\label{kgs}
\end{gather}
and $|D\rangle$ is the Dirichelet state and the explicit expression is in Appendix \ref{bstate}. It should be noted that since the mass does not go to zero at any finite time, the above state should is only valid in sufficiently long time limit and the correction due to the non-vanishing mass is $\mathcal{O}(e^{-\rho t})$.

\subsection{From squeezed states: CC state and gCC states}
We could start with specially prepared squeezed states so that after the quench, the states become CC states or gCC states. Here, we will consider only the simple case of sudden quench ($\rho\to\infty$). For our aim of creating a CC state or a gCC state, finite `$\rho$' quenches are an unnecessary complication.\\

We start with a squeezed state of `in' modes
\begin{eqnarray}
 |S\rangle = \exp\left(\sum_{k=-\infty}^{\infty}f(k)a^\dagger_{in,k}b^\dagger_{in,-k}\right)|0,in\rangle\
 \label{sqg}
\end{eqnarray}
where unlike $\gamma(k)$, $f(k)$ need not be an even function of $k$, but $|f(k)|^2$ is an even function of $k$.

It is easier to work with $|S\rangle$ as an operator relation. $|S\rangle$ can also be defined as
\begin{eqnarray}
 \label{sstate} \tilde{a}_{k}|S\rangle=\tilde{b}_{k}|S\rangle=0\
 \text{and} && \left\{\tilde{a}_k,\tilde{a}^\dagger_{k'}\right\}=\left\{\tilde{b}_{-k},\tilde{b}^\dagger_{-k'}\right\}=\delta(k-k')\
 \label{tacom}
\end{eqnarray}
where the new operators in terms of the out modes using (\ref{ain}) and (\ref{bdin}) are
\begin{align}
\tilde{a}_{k}&=\frac{1}{\sqrt{(1+|f(k)|^2)}}a_{in,k}-\frac{f(k)}{\sqrt{(1+|f(k)|^2)}}b^\dagger_{in,-k}\nonumber\\
&=A^*(k)a_{out,k}-\text{sgn}(k)B^*(k)b_{out,-k}\nonumber\\
\tilde{b}_{-k}&=\frac{1}{\sqrt{(1+|f(k)|^2)}}b_{in,-k}+\frac{f(k)}{\sqrt{(1+|f(k)|^2)}}a^\dagger_{in,k}\nonumber\\
&=A^*(k)b_{out,-k}+\text{sgn}(k)B^*(k)a^\dagger_{out,k}\
\label{tab}
\end{align}
where $A(k)$ and $B(k)$ are the Bogoliubov coefficients for the transformation from `{\it{tilde}}' operators to `out' operators and are given by
\begin{gather}
\label{AnB}A(k)=\frac{\alpha_+(k)-\text{sgn}(k)\beta_+^*(k)f^*(k)}{\sqrt{(1+|f(k)|^2)}},\quad B(k)=\frac{\beta_+(k)+\text{sgn}(k)\alpha_+^*(k)f^*(k)}{\sqrt{(1+|f(k)|^2)}}\\
|A(k)|^2+|B(k)|^2=1\
\label{AB1}
\end{gather}
Now using the BCH formula (\ref{BCHeqn}) from appendix (\ref{bch}),
\begin{gather}
\label{sqgo}|S\rangle=\exp\left\{-\sum_k \tilde{\kappa}(k)\left(a^\dagger_{out,k}a_{out,k}+b^\dagger_{out,k}b_{out,k}\right)\right\}|D\rangle\\
\text{where} \quad \tilde{\gamma}(k)=\frac{B^*(k)}{A^*(k)}, \quad\text{and}\quad \tilde{\kappa}(k)=-\frac{1}{2}\log(\tilde{\gamma}(k))\nonumber\
\end{gather}
For a CC state, i.e., so that $|S\rangle$ in eqn (\ref{sqgo}) is $e^{-\kappa_2 H}|D\rangle$, $f(k)$ should be tuned as
\begin{eqnarray}
 f(k)=\frac{\left(\sqrt{k^2+m^2}+m\right) \cosh (\kappa_2 k )-k \sinh (\kappa_2 k)}{\left(\sqrt{k^2+m^2}+m\right) \sinh (\kappa_2 k)+k\cosh (\kappa_2 k )}\
 \label{fcc}
\end{eqnarray}
Starting with
\begin{eqnarray}
 f(k)=\frac{k-k\, e^{2 \left| k\right|  \left(\kappa_2+\kappa_4 k^2\right)}+\text{sgn}(k)\left(\sqrt{k^2+m^2}+m\right)\left(e^{2 \left| k\right|  \left(\kappa_2+\kappa_4k^2\right)}+1\right)}{\left| k\right|  \left(e^{2 \left| k\right|  \left(\kappa_2 +\kappa_4 k^2 \right)}+1\right)+\left(\sqrt{k^2+m^2}+m\right) \left(e^{2 \left| k\right|  \left(\kappa_2+\kappa_4 k^2\right)}-1\right)}\
 \label{fgcc}
\end{eqnarray}
we get a gCC state of the form $e^{-\kappa_2 H-\kappa_4 W_4}|D\rangle$, where as mentioned earlier, $W_4$ is the conserved charge of the $\mathcal{W}_4$ current of free Dirac fermions{\footnote{For the action (\ref{action}), $H=\sum_k |k|(a^\dagger_k a_k+b^\dagger_k b_k)$ or $H=\int\frac{dk}{2\pi} |k|(a^\dagger_k a_k+b^\dagger_k b_k)$. $\mathcal{W}_4$ has been normalized so that $W_4=\sum_k |k|^3(a^\dagger_k a_k+b^\dagger_k b_k)$ or $W_4=\int \frac{dk}{(2\pi)^3}|k|^3(a^\dagger_k a_k+b^\dagger_k b_k)$ in the continuum limit.}. Note that $f(k)$ are odd functions of $k$.
For future reference, we can invert Eq (\ref{tab}) and we write down the `in' and `out' operators in terms of the `{\it{tilde}}' operators.
\begin{gather}
 \label{aintota}a_{in,k}=\frac{1}{\sqrt{(1+|f(k)|^2)}}\tilde{a}_k+\frac{f(k)}{\sqrt{(1+|f(k)|^2)}}\tilde{b}^\dagger_{-k}\\
 \label{bintotb} b^\dagger_{in,-k}=\frac{1}{\sqrt{(1+|f(k)|^2)}}\tilde{b}^\dagger_{-k}-\frac{f^*(k)}{\sqrt{(1+|f(k)|^2)}}\tilde{a}_k\\
 \label{aouttota} a_{out,k}=A(k)\tilde{a}_k+\text{sgn}(k)B^*(k)\tilde{b}^\dagger_{-k}\\
 \label{bdouttotbd}b^\dagger_{out,-k}=A^*(k)\tilde{b}^\dagger_{-k}-\text{sgn}(k)B(k)\,\tilde{a}_k\
\end{gather}
 
\section{\label{ed}Energy density}
In the post-quench theory, the occupation number is given by
\begin{eqnarray}
\hat{N}_k=a_k^\dagger a_k +b_k^\dagger b_k\ 
\end{eqnarray}
using the Bogoliubov transformations (\ref{aouttota}) and (\ref{bdouttotbd}) and definition of $|\tilde{0}\rangle$ in (\ref{sstate}), the expectation value of the occupation number is given by
\begin{eqnarray}
 N_k&=&\lim_{t\to\infty}\langle\tilde{0}|a_k^\dagger a_k +b_k^\dagger b_k|\tilde{0}\rangle\nonumber\\
 &=&B^*(k)B(k)\langle\tilde{0}|\tilde{b}_{-k}^\dagger\tilde{b}_{-k}|\tilde{0}\rangle+B^*(-k)B(-k)\langle\tilde{0}|\tilde{a}_{-k}^\dagger\tilde{a}_{-k}|\tilde{0}\rangle+...\nonumber\\
 &=& B^*(k)B(k)+B^*(-k)B(-k)\
\end{eqnarray}
The expression of $B(k)$ is given in (\ref{AnB}). For ground state, we have to use $f(k)=0$ in the expression of $B(k)$. So, energy density of the post-quench state is given by
\begin{eqnarray}
 E=\int_{-\infty}^\infty\frac{dk}{2\pi}|k|\left[B^*(k)B(k)+B^*(-k)B(-k)\right]\
\end{eqnarray}

\subsection*{Ground state quench}
For ground state quench, the occupation number is given by
\begin{eqnarray}
 N_k&=&\lim_{t\to\infty}\langle 0,in|\hat{N}_k|0,in\rangle=|\beta_+(k)|^2+|\beta_+(-k)|^2\nonumber\
\end{eqnarray}
Since $\alpha_+(k)$ and $\beta_+(k)$ are even functions of $k$. Using (\ref{ab1}), (\ref{gg}) and (\ref{forbch}), we have
\begin{eqnarray}
|\beta_+(k)|^2=\frac{|\gamma(k)|^2}{1+|\gamma(k)|^2},\qquad |\gamma(k)|^2=e^{-4\kappa(k)}\
\end{eqnarray}
Hence, the occupation number in the ground state in the asymptotically long time limit is given by
\begin{eqnarray}
 N_k=\frac{2}{e^{4\kappa(k)}+1}\
\end{eqnarray}
This is the occupation number in a GGE defined as
\begin{eqnarray}
\text{Tr}\,e^{-\sum_k 4\kappa(k)\hat{N}_k}=\text{Tr} e^{- 4\kappa_2 H-4\kappa_4 W_4-\kappa_6 W_6-\cdots}\
\end{eqnarray}
where the $\kappa$'s are given in (\ref{kgs}).
Using the expressions of $\beta_+(k)$ from (\ref{beta}), the explicit expression of the occupation number is 
\begin{eqnarray}
 N_k&=&\text{csch}\left(\frac{\pi  k}{\rho }\right) \left(\cosh \left(\frac{\pi  m}{\rho }\right)-\cosh \left(\frac{\pi  \left(k-\sqrt{k^2+m^2}\right)}{\rho }\right)\right) \text{csch}\left(\frac{\pi  \sqrt{k^2+m^2}}{\rho }\right)\nonumber\\
 &\xrightarrow{\rho\to\infty}&1-\frac{k}{\sqrt{k^2+m^2}}\nonumber\
\end{eqnarray}
It is interesting that in $m\to\infty$ limit, $N_k\to1$, not 2. This is because $\lim_{\rho\to\infty}|\alpha_+(k)|^2=1/2$ and we have the constraint $|\alpha_+(k)|^2+|\beta_+(k)|^2=1$.\\
For arbitrary $\rho$, the energy density cannot be calculated in closed form. In the sudden limit $\rho\to\infty$, the energy density diverges as $\log(\Lambda)$ where $\Lambda$ is the UV cutoff. Hence, all other $W$ charges also diverge in the sudden limit. Hence, naively taking $\rho\to\infty$ produce a non-renormalizable state. So, the sudden limit has to be taken as in \cite{Mandal:2015kxi} where $m/\Lambda\to0$ while $m/\rho\to\epsilon^+$. Simply put, the quench rate parameter $\rho$ should be much small than the UV cut-off.

\subsection*{Squeezed state quench: CC and gCC states}
For CC state given by (\ref{fcc}), the expectation value of occupation number is given by
\begin{eqnarray}
 N_k=\frac{2}{1+e^{4 \kappa_2\left|k\right|}}\
\end{eqnarray}
This is the occupation number of fermions in a thermal ensemble of temperature $1/\beta=1/4\kappa_2$. The enengy density is
\begin{eqnarray}
 E=\int_{-\infty}^{\infty}\frac{dk}{2\pi}N_k=\frac{\pi }{96 \kappa_2 ^2}\
\end{eqnarray}
Similarly, for gCC state given by (\ref{fgcc}), the expectation value of occupation number is given by
\begin{eqnarray}
 N_k=\langle gCC|\hat{N}_k|gCC\rangle=\frac{2}{1+e^{4 \kappa_2\left|k\right|+4\kappa_4\left|k\right|^3}}\
\end{eqnarray}
This is same as the occupation number of fermions in a generalised Gibbs ensemble of temperature $1/\beta=4\kappa_2$ and chemical potential $\mu_4=4\kappa_4$ of $W_4$ charge. The enengy density cannot be calculated in closed form.

\section{\label{crf}Correlation functions}
Since our theory is a free theory, all the observables can be explicitly calculated. In the following subsections we calculate $\langle\psi^\dagger(r,t)\psi(0,t)\rangle$ correlation functions for the three different states obtained above. The quench process cannot differentiate between holomorphic dof(`left-movers') and anti-holomorphic dof(`right-movers'), so $\langle\bar{\psi}^\dagger(0,t)\bar{\psi}(r,t)\rangle$ is equal to $\langle\psi^\dagger(r,t)\psi(0,t)\rangle$ and they are time independent quantities.{\footnote{A simple reason why these quantities are time independent is the fact that they are holomorphic-holomorphic and antiholomorphic-antiholomorphic quantities and they cannot `see' the presence of the boundary state $|D\rangle$. They are already thermalized/equilibrated.}} We also calculated $\langle\bar{\psi}^\dagger(r,t)\psi(0,t)\rangle$ which has non-trivial time-dependence. Also as expected, $-\langle\psi^\dagger(0,t)\bar{\psi}(r,t)\rangle$ is the complex conjugate of $\langle\bar{\psi}^\dagger(r,t)\psi(0,t)\rangle$. Since, we are calculating equal-time correllation functions, so for example for $\langle\psi^\dagger(r,t)\psi(0,t)\rangle$, we would rather be calculating $\frac{1}{2}\langle\psi^\dagger(r,t)\psi(0,t)-\psi(0,t)\psi^\dagger(r,t)\rangle$.\\
Using the Bogoluibov transformations (\ref{aouttota}) and (\ref{bdouttotbd}) in the chiral mode expansions (\ref{crep}) and (\ref{crepb})  we get 
\begin{align}
\label{psioutt}\psi(w)&=\int_0^\infty \frac{dk}{2\pi} \Big[A(k) \tilde{a}_{k}e^{-ikw}+\text{sgn}(k)B^*(k)\tilde{b}^\dagger_{-k}e^{-ikw}+A^*(-k)\tilde{b}^\dagger_{k}e^{ikw}+\text{sgn}(k)B(-k)\tilde{a}_{-k}e^{ikw}\Big]\\
\bar{\psi}(\bar{w})&=\int^{\infty}_0 \frac{dk}{2\pi} \Big[A(-k) \tilde{a}_{-k}e^{-ik\bar{w}}-\text{sgn}(k)B^*(-k)\tilde{b}^\dagger_{k}e^{-ik\bar{w}}-A^*(k)\tilde{b}^\dagger_{-k}e^{ik\bar{w}}+\text{sgn}(k)B(k)\tilde{a}_{k}e^{ik\bar{w}}\Big]\
\label{bpsioutt}
\end{align}
where $w=t-x$ and $\bar{w}=t+x$. For the ground state quench, $f(k)=0,\,\tilde{a}_{k}=a_{in,k},\,\tilde{b}=b_{in,k}$ and $|\tilde{0}\rangle=|0,in\rangle$.\\
For a general $f(k)$ corresponding to some $|\tilde{0}\rangle$, the correlation functions are
 \begin{eqnarray}
 \label{psidpsiS}\langle\tilde{0}|\psi^\dagger(0,t)\psi(r,t)|\tilde{0}\rangle &=& \frac{1}{2}\int_{0}^{\infty}\frac{dk}{2\pi}\left[(2|B(k)|^2-1)e^{ikr}-(2|B(-k)|^2-1)e^{-ikr}\right]\\
 \label{bpsidpsiS}\langle\tilde{0}|\bar{\psi}^\dagger(0,t)\psi(r,t)|\tilde{0}\rangle&=& -\int_0^{\infty}\frac{dk}{2\pi}\left[\text{sgn}(k)A^*(-k)B(-k)e^{ik(2t-r)}+\text{sgn}(k)A(k)B^*(k)e^{-ik(2t-r)}\right]\nonumber\\
 \end{eqnarray}
where we have used (\ref{AB1}) to write $A(k)$ in terms of $B(k)$ in the first equation.\\
\subsection{Ground state quench}
Taking careful limit, for ground state quench, we have
\begin{eqnarray}
 \langle 0,in|\psi^\dagger(0,t)\psi(r,t)|0,in\rangle &&= -\frac{1}{2}\int_{-\infty}^{\infty}\frac{dk}{2\pi}\frac{\left| k\right| }{\sqrt{k^2+m^2}}\nonumber\\
 &&=\frac{1}{4} m \left[\pmb{L}_{-1}(m r)-I_1(m r)\right]\nonumber\\
 &&\xrightarrow{m\to\infty}\frac{1}{2 \pi  m r^2}+\frac{3}{2 \pi  m^3 r^4}+O\left(\frac{1}{m^4}\right)\\
 \langle\tilde{0}|\bar{\psi}^\dagger(0,t)\psi(r,t)|\tilde{0}\rangle&&=\int_0^{\infty}\frac{dk}{2\pi}\,\frac{i \,\text{sgn}(k) \,m\sin (k (2 t-r))}{\sqrt{k^2+m^2}}\nonumber\\
 &&=-\frac{i m}{4} \, \left[\text{sgn}(r-2 t) I_0(m (r-2 t))-\pmb{L}_0(m (r-2 t))\right]\nonumber\\
 &&\xrightarrow[t>r/2]{m\to\infty}\frac{i}{2 \pi  (2 t-r)}+\frac{i}{2 \pi  m^2 (2 t-r)^3}+O\left(\frac{1}{m^4}\right)\
\end{eqnarray}
where $I_\nu(x)$ is Modified Bessel Function of the First Kind and $\pmb{L}_\nu(x)$ is Modified Struve Function.
\subsection{Quenched squeezed state - CC state}
For CC state, all the calculations are done in $|S_{CC}\rangle$ defined as the state (\ref{sqg}) with the expression of $f(k)$ given in (\ref{fcc}).
\begin{eqnarray}
 \langle CC|\psi^\dagger(0,t)\psi(r,t)|CC\rangle&=&-i\int_{0}^{\infty}\frac{dk}{2\pi}\,\tanh (2\kappa_2 |k|)\sin(kr)\\
 \label{thermform}&=&-i\int_{0}^{\infty}\frac{dk}{2\pi}\,\sin(kr)\left[\frac{1}{e^{4\kappa_2 |k|}+1}-\frac{1}{2}\right]\\
 &=& -\frac{i\, \text{csch}\left(\frac{\pi  r}{4 \kappa _2}\right)}{8 \kappa _2}\\
\label{ccr=0} \langle CC|\bar{\psi}^\dagger(0,t)\psi(r,t)|CC\rangle&=&-i\int_0^{\infty}\frac{dk}{2\pi}\,\text{sech}(2 k \kappa_2 ) \cos (k(2 t-r))\\
 &=&-\frac{i\, \text{sech}\left(\frac{\pi  (2t-r)}{4 \kappa_2 }\right)}{8 \kappa_2 }\
\end{eqnarray}
These are exactly what have been calculated using BCFT techniques \cite{Calabrese:2006quench}. It is evident from (\ref{thermform}) that $\psi^\dagger\psi$ expectation value is already the thermal expectation value at temperature $T=1/\beta =1/(4\kappa_2)$, i.e., it is already thermalized.
\subsection{Quenched squeezed state - gCC state with $W_4$}
Similarly, for gCC state, all the calculations are done in $|S_{fCC}\rangle$ defined as the state (\ref{sqg}) with the expression of $f(k)$ given in (\ref{fgcc}).
\begin{eqnarray}
 \langle\psi^\dagger(0,t)\psi(r,t)\rangle_{gCC}&=&-i\int_{0}^{\infty}\frac{dk}{2\pi}\,\tanh\left(2\kappa_2 |k| +2\kappa_4 |k|^3\right)\,\sin(kr)\\
\label{thermform2} &=&-i\int_{0}^{\infty}\frac{dk}{2\pi}\,\sin(kr)\left[\frac{1}{e^{4\kappa_2 |k|+4\kappa_4 |k|^3}+1}-\frac{1}{2}\right]\\
 \langle\bar{\psi}^\dagger(r,t)\psi(0,t)\rangle_{gCC}&=&-i\int_0^{\infty}\frac{dk}{2\pi}\,\text{sech}(2 \kappa_2 k+2\kappa_4 k^3 ) \cos (k(2 t-r))\
 \end{eqnarray}
Again, it is evident from (\ref{thermform2}) that $\psi^\dagger\psi$ expectation value is already thermalized into the expectation value in a GGE with $T=1/\beta =1/4\kappa_2$ and $\mu=4\kappa_4$. A possible way of evaluating these integrals (which yield no closed form answer) is via the residue theorem. The integrands in both cases, have poles at the solutions of $2\kappa_2 k +2\kappa_4 k^3= \frac{2n+1}{2} i\pi$, where $n\in\mathbb{Z}$. These poles and their residues have been treated in detail in \cite{Mandal:2015kxi}. The sum of residues is still an infinite sum which cannot be performed. In the perturbative regime ($\kappa_4/\kappa_2^3<<1$), we see that our correlators match the general form presented in \cite{Mandal:2015jla} with $h=1/2$ as expected.

\section{Entanglement entropy and mutual information in CC state}
\label{secEE_CC}
In this section, we will take a closer look at entanglement dynamics in CC state.
For simplicity, first we will consider a finite single interval or subsystem A, with its endpoints at $(w_1,\bar{w}_1)$ and $(w_2,\bar{w}_2)$ in light-cone coordinates, or $(0,t)$ and $(r,t)$ in space and time coordinates. Using the replica trick (\cite{Holzhey:1994we}, \cite{Calabrese:2004eu}), the $n\textsuperscript{th}$ R\'enyi entropy $S_n(A)$ of the interval is given by the logarithm of the expectation value of twist and antitwist operators inserted at the end-points.
\begin{eqnarray}
\label{sna} S_n(A)=\frac{1}{1-n}\log\langle\Psi(t)|\mathcal{T}_{n}(w_1,\bar{w}_1)\tilde{\mathcal{T}}_{n}(w_2,\bar{w}_2)|\Psi(t)\rangle\
\end{eqnarray}
The entanglement entropy(EE) $S_{A}$ is given by $\lim_{n\to1}S_n(A)$.
We can diagonalize the twist operators and write them as products of twist fields. Hence,
\begin{eqnarray}
\label{tntkn} \mathcal{T}_{n}(w,\bar{w})=\prod_{k=-(n-1)/2}^{k=(n-1)/2}\mathcal{T}_{k,n}(w,\bar{w}), \quad  \tilde{\mathcal{T}}_{n}(w,\bar{w})=\prod_{k=-(n-1)/2}^{k=(n-1)/2}\tilde{\mathcal{T}}_{k,n}(w,\bar{w})\
\end{eqnarray}
In CC state, in Heisenberg picture, the quantity of our interest is
\begin{eqnarray}
\label{tttf} Z_{k}=\langle D_{f}|e^{-\kappa_2 H_{f}}\mathcal{T}_{k,n}(0,t)\tilde{\mathcal{T}}_{k,n}(r,t)e^{-\kappa_2 H_{f}}|D_{f}\rangle\
\end{eqnarray}
The subscript `f' means we are working in the fermionic theory and the subscript `b' would mean we are working in the bosonic theory. To find the exact expression of the entanglement entropy of a spatial region in our free fermionic CFT, we will use the method of bosonization described in \cite{Casini:2005rm}. Moreover, as shown in Appendix(\ref{bbs}), Dirichlet state $|D_{f}\rangle$ in fermionic theory corresponds to a Dirichlet state in the bosonic theory $|D_{b}\rangle$ and $H_{f}$ corresponds to $H_{b}$. So, we get
\begin{eqnarray}
\label{tttf} Z_{k}&=&\langle D_{b}|e^{-\kappa_2 H_{b}}e^{i\sqrt{4\pi}\frac{k}{n}\left(\phi(0,t)-\phi(r,t)\right)}e^{-\kappa_2 H_{b}}|D_{b}\rangle\
\end{eqnarray}
This is a free scalar theory in a strip geometry with Dirichlet boundary conditions and operator insertions at $(0,t)$ and $(r,t)$. The path integration can be easily performed and the result gives
\begin{eqnarray}
 \log\left[Z_{k}\right]&=& -4\pi\,\frac{k^2}{n^2}\left(\langle\phi(0,t)\phi(0,t)\rangle-\langle\phi(0,t)\phi(r,t)\rangle\right)\
\end{eqnarray}
\noindent The $n\textsuperscript{th}$ R\'enyi entropy of interval A is given by
\begin{eqnarray}
  S_{nA}(t)&=&-4\pi\,\frac{1}{1-n}\sum_{k=-(n-1)/2}^{k=(n-1)/2}\frac{k^2}{n^2}\left(\langle\phi(0,t)\phi(0,t)\rangle-\langle\phi(0,t)\phi(r,t)\right.\nonumber\\
  &&\hspace{5cm} \left.-\langle\phi(r,t)\phi(0,t)\rangle+\langle\phi(r,t)\phi(r,t)\rangle\right)\nonumber\\
 \label{RE} &=& 4\pi\,\frac{n+1}{6 n}\,\left(\langle\phi(0,t)\phi(0,t)\rangle-\langle\phi(0,t)\phi(r,t)\rangle\right)\
\end{eqnarray}
Taking $n\to1$ limit, we get the entanglement entropy,
\begin{eqnarray}
\label{EE} S_{A}(t)=4\pi\,\frac{1}{3}\,\left(\langle\phi(0,t)\phi(0,t)\rangle-\langle\phi(0,t)\phi(r,t)\rangle\right)\
\end{eqnarray}
The generalization to the case for subsystem $A_q$ consists of $q$ disjoint intervals is straightforward. Consider the left end of the intervals to be at $u_i$ and right end of the intervals to be at $v_i$ where $i$ goes from $1$ to $q$. After the path integration, the $n\textsuperscript{th}$ R\'enyi entropy and the entanglement entropy are given by
\begin{eqnarray}
S_{nA_q}(t) = 4\pi\,\frac{n+1}{12 n}\sum_{i,j=1}^q\left(\langle\phi(u_i,t)\phi(u_j,t)\rangle+\langle\phi(v_i,t)\phi(v_j,t)\rangle-2\langle\phi(u_i,t)\phi(v_j,t)\rangle\right)\\
\label{CC_RE_multi}
\label{CC_EE_multi} S_{A_q}(t)=4\pi\,\frac{1}{6}\,\sum_{i,j=1}^q\left(\langle\phi(u_i,t)\phi(u_j,t)\rangle+\langle\phi(v_i,t)\phi(v_j,t)\rangle-2\langle\phi(u_i,t)\phi(v_j,t)\rangle\right)\
\end{eqnarray}
The coincident terms will give rise to the familiar UV divergent pieces of entanglement entropy in quantum field theory. We will concentrate on the entanglement entropy since the different $n\textsuperscript{th}$ R\'enyi entropy differs from entanglement entropy by a simple overall multiplicative factor.

\gap3
\noindent{\small{\bf{Remark on winding number:}} While the free boson considered in \cite{Mandal:2015kxi} is the uncompactified free boson, the boson in (\ref{tttf}) is a compactified free boson. So, Hamiltonian of the compactifed boson has zero mode terms but the winding number is not important for our analysis.
In the large system size limit($L\to\infty$), the zero modes vanished.
Even if we are taking the limiting case of a finite size system, the zero momentum modes do not play any role in our calculation. Using the mode expansion of the boson $\phi(w,\bar{w})=\varphi(w)+\bar{\varphi}(\bar{w})$ in \cite{Senechal:1999us},
\begin{eqnarray}
\varphi(w)&=&Q+\frac{P}{2L}\,w +\sum_{n>0}\frac{1}{\sqrt{4\pi n}}\left(d_n e^{-inw}+d^{\dagger}_ne^{inw}\right)\\
\bar{\varphi}(\bar{w})&=&\bar{Q}+\frac{\bar{P}}{2L}\,\bar{w}+\sum_{n>0}\frac{1}{\sqrt{4\pi n}}\left(d_{-n} e^{-in\bar{w}}+d^{\dagger}_{-n}e^{in\bar{w}}\right)\
\end{eqnarray}
First, $Q$ and $\bar{Q}$ are cancelled identically in (\ref{tttf}).
Moreover, by bosonization formulae \cite{Senechal:1999us, vonDelft:1998pk},
\begin{align}
 P&=\sqrt{4\pi}N_f & \bar{P}&=\sqrt{4\pi}\bar{N}_f\\
 N_f&=J_0=-\sum_{k=0}^{\infty}\left[a^\dagger_ka_k-b^\dagger_kb_k\right] & \bar{N}_f&=\bar{J}_0=-\sum_{k=0}^{\infty}\left[a^\dagger_{-k}a_{-k}-b^\dagger_{-k}b_{-k}\right]\
\end{align}
 But for our particular CC state, from (\ref{JbJ0}), $N_f|CC_f\rangle = 0$ and $\bar{N}_f|CC_f\rangle = 0$. Now, $P$ and $\bar{P}$ commute with all the other bosonic creation and annihilation operators of non-zero momentum, hence they don't play any role in the calculation of (\ref{tttf}).
 If we still keep the system size finite, the winding number would be important to interpret the stationary limit as a thermal ensemble. But we must take the $L\to\infty$ limit, if we want to examine the stationary limit. In other words, $L$ is the largest length scale in our theory and time $t<<L$.}
\gap3

The bosonic propagator in CC state has been calculated in \cite{Mandal:2015kxi}. It is given by
\begin{eqnarray}
\label{phiprop}\langle CC|\phi(0,t)\phi(r,t)|CC\rangle=-\frac{1}{8\pi}\log \left(\frac{2\sinh ^2\left(\frac{\pi  r}{4 \kappa_2 }\right)}{\cosh \left(\frac{\pi  r}{2 \kappa_2 }\right)+\cosh \left(\frac{\pi  t}{\kappa_2 }\right)}\right)\
\end{eqnarray}
The entanglement entropy in CC state of single interval A of length $r$ is given by
\begin{eqnarray}
\label{CC_EE_one} S_{A}(t)=\frac{1}{3}\,\left[\frac{1}{2}\log \left(\frac{\sinh ^2\left(\frac{\pi  r}{4 \kappa_2 }\right)\left(1+\cosh \left(\frac{\pi  t}{\kappa_2 }\right)\right)}{\cosh \left(\frac{\pi  r}{2 \kappa_2 }\right)+\cosh \left(\frac{\pi  t}{\kappa_2 }\right)}\right) -\lim_{\epsilon\to 0+}
\log \left(\frac{\pi \epsilon}{4 \kappa_2}\right)\right]\
\end{eqnarray}

\begin{figure}
  \centering
  \begin{subfigure}{.5\textwidth}
  \includegraphics[width=.9\linewidth]{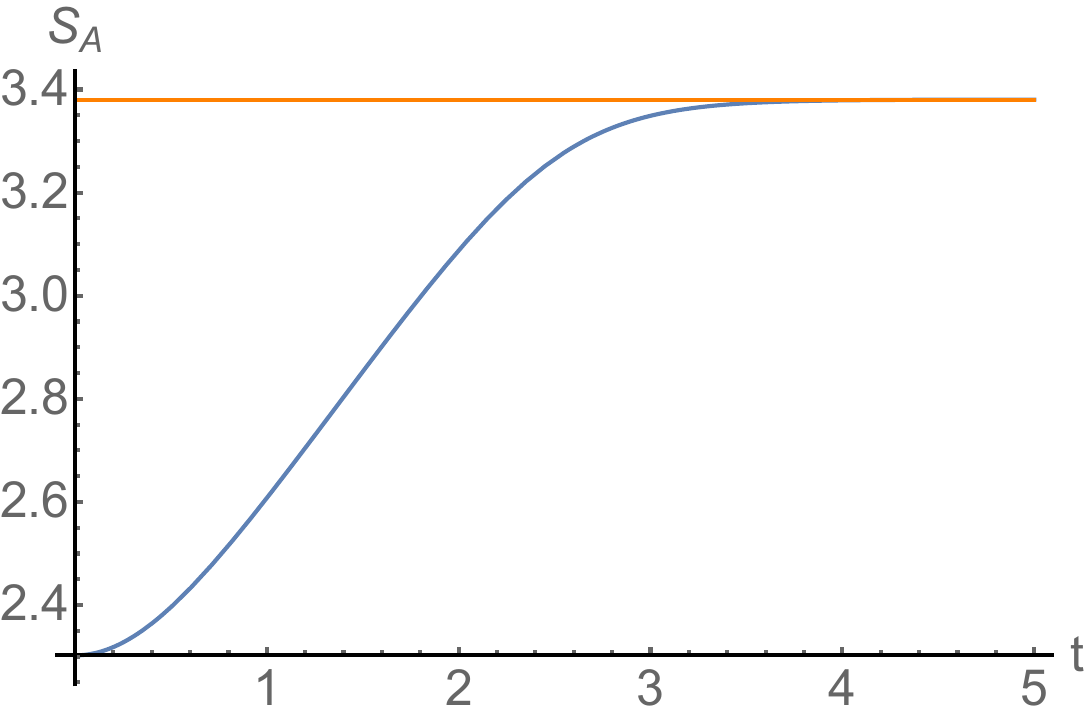}
  \caption{\small Low effective temperature, $\kappa_2=1$}
  \label{fig:CC_EE_low_temp}
\end{subfigure}%
\begin{subfigure}{.5\textwidth}
  \centering
  \includegraphics[width=.9\linewidth]{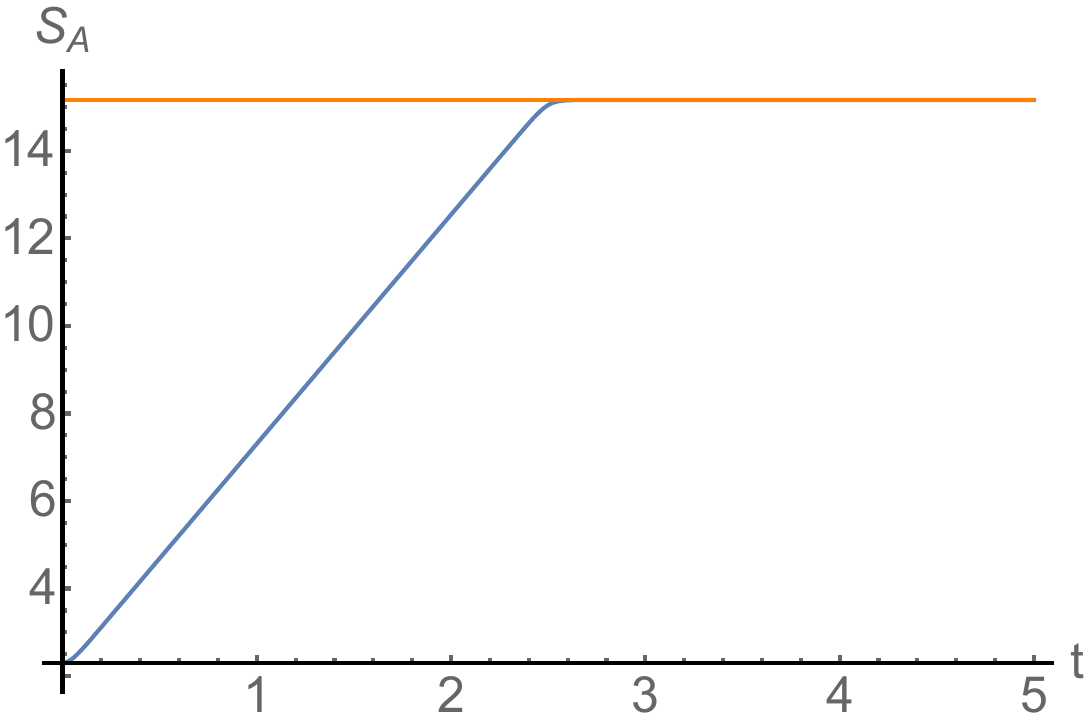}
  \caption{\small High effective temperature, $\kappa_2=.1$}
  \label{fig:CC_EE_high_temp}
\end{subfigure}
\caption{\small Entanglement entropy growth of an interval(r=5) in CC state.}
\label{fig:CC_SA}
\end{figure}

Figure \ref{fig:CC_SA} consists of plots of $S_A(t)$ as a function of time $t$ for different values of the effective temperature. For a large subsystem (compare to the effective temperature), the entanglement growth is linear which is well explained by the quasiparticle picture described in \ref{sec:intro}. The above simple closed form expression has many interesting features. We enumerate these features below.
\begin{enumerate}
\item {\bf Stationary limit:} Taking the stationary limit $t\to\infty$ gives the entanglement entropy of A in a thermal ensemble at temperature $T=1/\beta=1/(4\kappa_2)$.
\begin{eqnarray}
 S_{A}&=&\frac{1}{3}\,\left[\log \left(\sinh \left(\frac{\pi  r}{4 \kappa_2 }\right)\right)-\lim_{\epsilon\to 0+}\log \left(\frac{\pi\epsilon}{4 \kappa_2}\right)\right]\
 \label{SA_therm}
\end{eqnarray}
This exactly matches the thermal value which has been calculated using CFT techniques \cite{Calabrese:2004eu}. It is fixed only by the temperature, the central charge of the CFT($c=1$ for a Dirac fermion). Taking the high temperature limit $\kappa_2\to 0$, we get the extensive thermal entropy formula $S_{therm}=\frac{1}{3}\frac{\pi r}{\beta}$.

\item {\bf Monotonic growth:} Besides thermalization, the most interesting aspect of (\ref{CC_EE_one}) is that the entanglement entropy of a single interval grows monotonically. The first derivative of $S_A$ w.r.t. time is
\begin{eqnarray}
 \left<\frac{\partial S_A}{\partial t}\right>_{CC} &=& \frac{\pi  \sinh ^2\left(\frac{\pi  r}{4 \kappa_2 }\right) \tanh \left(\frac{\pi  t}{2 \kappa_2 }\right)}{3 \kappa_2 \left[ \cosh \left(\frac{\pi  r}{2 \kappa_2 }\right)+  \cosh \left(\frac{\pi  t}{\kappa_2 }\right)\right]}\\
 \label{scctan} &=&\frac{\pi}{12 \kappa_2 }  \left[2 \tanh \left(\frac{\pi  t}{2 \kappa_2 }\right)-\tanh \left(\frac{\pi  (r+2 t)}{4 \kappa_2 }\right)+\tanh \left(\frac{\pi  (r-2 t)}{4 \kappa_2 }\right)\right]\
\end{eqnarray}
From the first expression, as a function of time $t>0$, it is clear that there are no finite zero. Hence, the EE growth of CC state is always monotonically increasing. Also note that in the high effective temperature limit $\kappa_2\to 0$, the approach to thermal value is sharper. In the limiting case, from the second expression, it is clear that the thermalization time is 
\begin{equation}
\label{thermtime}t=\frac{r}{2}\ 
\end{equation}
which has also been calculated using BCFT techniques in \cite{Calabrese:2005in}.

\item {\bf Early time growth:} The entanglement entropy has a quadratic growth at early times. The expansion of (\ref{CC_EE_one}) at early time is given by
\begin{eqnarray}
S_A(t)&=&\log \left(\frac{2 \sinh ^2\left(\frac{\pi  r}{4 \kappa }\right)}{\cosh \left(\frac{\pi  r}{2 \kappa }\right)+1}\right)+\frac{t^2 \left(\pi ^2 \cosh \left(\frac{\pi  r}{2 \kappa }\right)-\pi ^2\right)}{\left(4 \kappa ^2\right) \left(\cosh \left(\frac{\pi  r}{2 \kappa }\right)+1\right)}\nonumber\\
&&\qquad +\frac{t^4 \left(-\pi ^4 \cosh ^2\left(\frac{\pi  r}{2 \kappa }\right)-\left(6 \pi ^4\right) \cosh \left(\frac{\pi  r}{2 \kappa }\right)+7 \pi ^4\right)}{\left(96 \kappa ^4\right) \left(\cosh \left(\frac{\pi  r}{2 \kappa }\right)+1\right)^2}+\mathcal{O}\left(t^6\right)\
\end{eqnarray}
In the large $r$ limit, the quadratic term is $\pi^2t^2/(4\kappa_2^2)$.
It is interesting that similar quadratic growth, although with a different prefactor, has also been found in holographic set-up where the entanglement growth in $AdS_3$ Vaidya metric is calculated \cite{Liu:2013qca}. But note that the exact holographic dual of CC states (of holographic CFTs) are the single sided black holes with end-of-world branes \cite{Hartman:2013qma}.

\end{enumerate}

In the limit $r\to 0$, one should be careful and take into account the UV term also. Otherwise, the first term alone in (\ref{CC_EE_one}) and (\ref{SA_therm}) will become negative.

Figure \ref{fig:CC_SAmultiple} are the plots of the entanglement entropy of subsystems consisting of different number of disjoint intervals at different effective temperature. For simplicity, we have considered intervals of same length $r=5$ separated by intervals of the same length $l=r$. Qualitatively at low effective temperature and quantitatively at high effective temperature, the entanglement dynamics is well explained by the quasiparticle picture.

\begin{figure}
\centering
\begin{subfigure}{.5\textwidth}
  \centering
  \includegraphics[width=.9\linewidth]{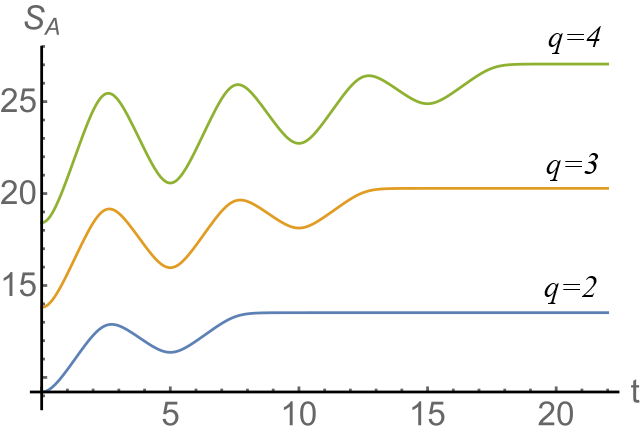}
  \caption{\small Low effective temperature, $\kappa_2=1$}
  \label{fig:CC_SA_kappa1}
\end{subfigure}%
\begin{subfigure}{.5\textwidth}
  \centering
  \includegraphics[width=.9\linewidth]{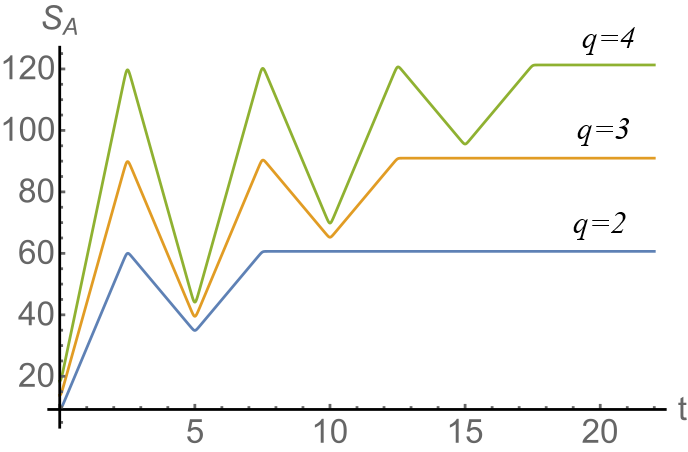}
  \caption{\small High effective temperature, $\kappa_2=.1$}
  \label{fig:CC_SA_kappa01}
\end{subfigure}
\caption{\small Entanglement entropy growth  in CC state of subsystems consisting of multiple intervals ($q=$ number of intervals). For simplicity we choose interval-size $(r=5)$ separated by distance $l=r$.}
\label{fig:CC_SAmultiple}
\end{figure}

Mutual information of two disjoint subsystem is defined as
\begin{equation}
\label{MI_def}
I(A,B)=S_A+S_B-S_{A\bigcup B}\
\end{equation}
It measures the quantum\footnote{We consider only pure states so the nature of the correlation is quantum mechanical. In case of thermal states, there would be thermal correlation also at short distance.} correlation between the two disjoint subsystems. In CC state, it is easy to calculate mutual information between two disjoint intervals using (\ref{CC_EE_multi}), (\ref{phiprop}) and (\ref{CC_EE_one}). Figure \ref{fig:CC_MI} are plots of the mutual information between two disjoint intervals of length $r=5$ separated by a distance $l=5$. The MI starts growing from $t\sim l/2$ and becomes negligible again from $t\sim r+l/2$. This is again well explained by the quasiparticle picture.
\begin{figure}
  \centering
  \begin{subfigure}{.5\textwidth}
  \includegraphics[width=.9\linewidth]{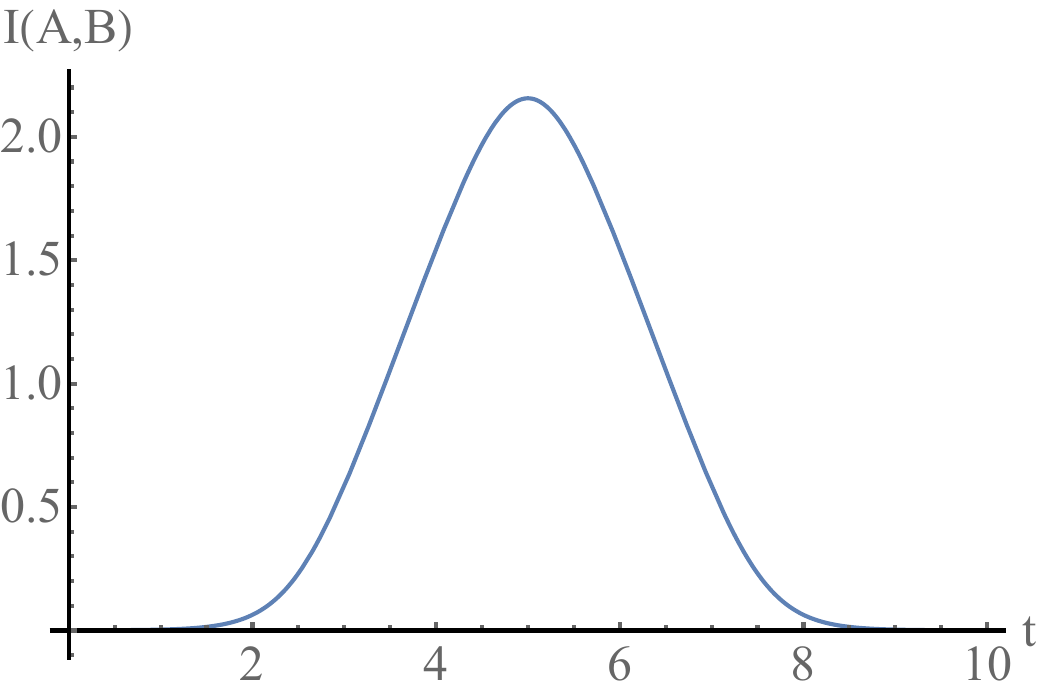}
  \caption{\small Low effective temperature, $\kappa_2=1$}
  \label{fig:CC_MI_low_temp}
\end{subfigure}%
\begin{subfigure}{.5\textwidth}
  \centering
  \includegraphics[width=.9\linewidth]{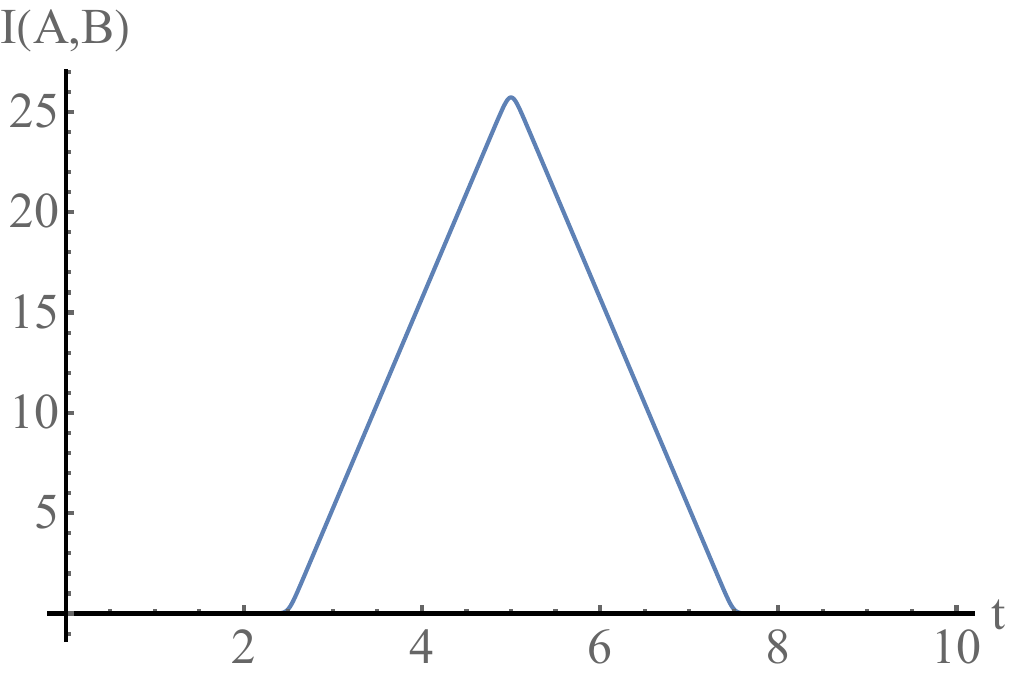}
  \caption{\small High effective temperature, $\kappa_2=.1$}
  \label{fig:CC_MI_high_temp}
\end{subfigure}
\caption{\small Mutual information of two disjoint intervals of each of length $r=5$ separated by a distance $l=5$.}
\label{fig:CC_MI}
\end{figure}

It would be interesting to check the monotonicity of EE growth in gCC states. Unfortunately, even for the free fermions with explicit twist operators, the entanglement entropy in gCC state with $W_4$ charge cannot be explicitly calculated. The bilinear fermionic $\mathcal{W}_4(w)$ current when bosonized gives $\phi^4$ terms \cite{Pope:1991ig}, so the bosonized theory is an interacting theory. It is an interacting theory in the sense that the path integral on the strip as in (\ref{tttf}) would be that of an interacting theory. Note that the time evolution operator is still the Hamiltonian of a free theory.

\section{\label{secEE_gCC}Non-Monotonic EE Growth and Dynamical Phase Transition}
Although we could not calculate EE in gCC state with $W_4$ charge of the fermionic bilinear $\mathcal{W}_4$ current, we can still calculate entanglement entropy explicitly with the fermionic charge corresponding to the bilinear bosonic charge $W_4(w)=\sum_k |k|^3 d^\dagger_kd_k$, where $d^\dagger_k$ and $d_k$ are the bosonic annihilation and creation operators. As mentioned above, the zero modes do not play any role. Refermionization of the bosonic bilinear $\mathcal{W}_4$ is done in Appendix \ref{refW4}.\footnote{We would like to thank Justin David for informing us that this refermionization could be done in principle using U(1) currents and it has not been done anywhere.} So, the fermionic state that we are considering is
\begin{eqnarray}
 |\Psi\rangle=\text{e}^{-\kappa_2 H_f-\kappa_4 \tilde{W}_4}|D_f\rangle\
\end{eqnarray}
where the expression for $\tilde{W}_4$ is given in (\ref{newW4ch}).

Again, the R\'enyi and entanglement entropies are given by the expression (\ref{RE}) and (\ref{EE}). The scalar propagator with the bosonic $W_4$ charge has also been calculated in \cite{Mandal:2015kxi}.
\begin{eqnarray}
 \langle\phi(0,t)\phi(r,t)\rangle=\int_{-\infty}^{\infty}\frac{dk}{8\pi}\frac{e^{ikr}}{k} \left[\coth \left(2 k \left(\kappa_2 +\kappa_4  k^2\right)\right)-\cos (2 k t) \csch\left(2 k \left(\kappa_2 +\kappa_4  k^2\right)\right)\right]\
\end{eqnarray}
The momentum integral cannot be done explicitly. But we still can plot the entanglement entropy numerically. Figure (\ref{gCC_EE}) are the plots of EE growth with `small' and `large' values of $\kappa_4$. As expected, the entanglement entropy reaches an equilibrium quickly.

\begin{figure}
  \centering
  \begin{subfigure}{.5\textwidth}
  \includegraphics[width=.9\linewidth]{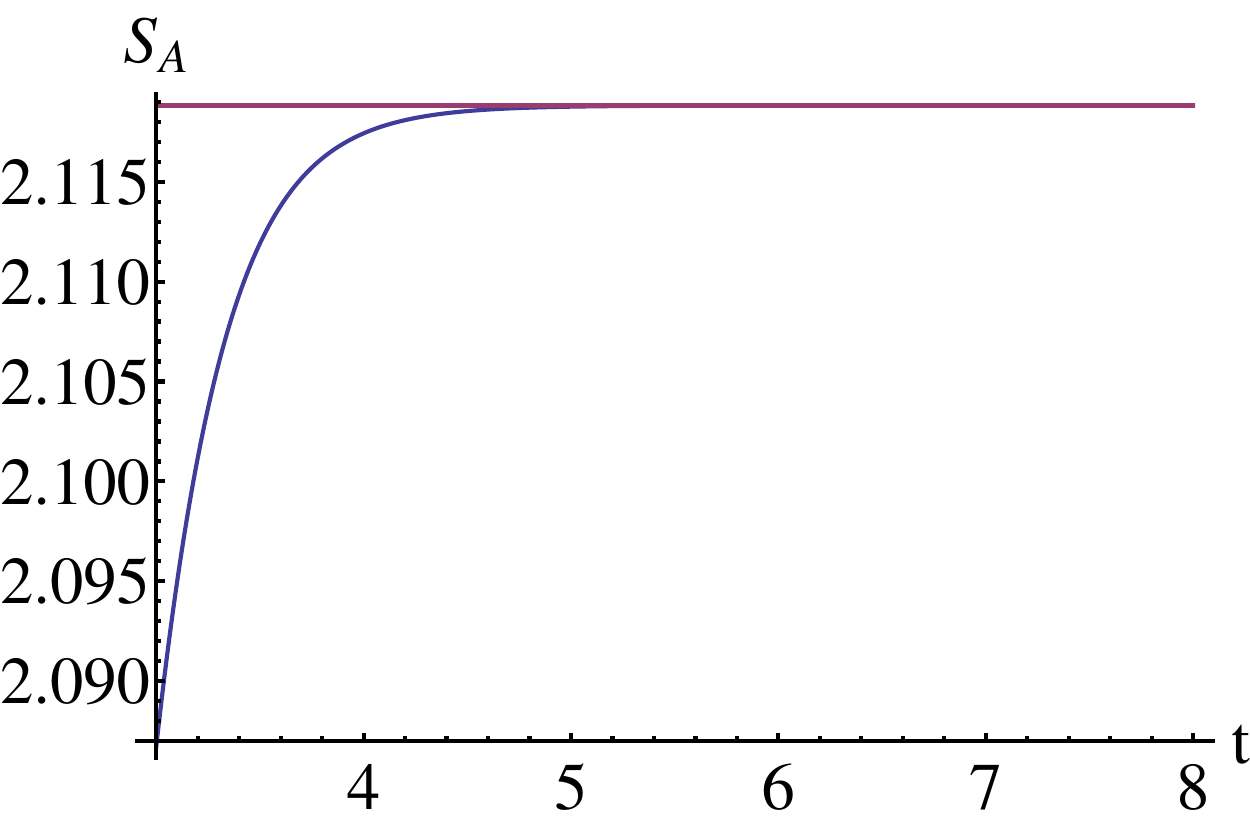}
  \caption{\small Monotonic behaviour, $\kappa_4=0.01$}
\end{subfigure}%
\begin{subfigure}{.5\textwidth}
  \centering
  \includegraphics[width=.9\linewidth]{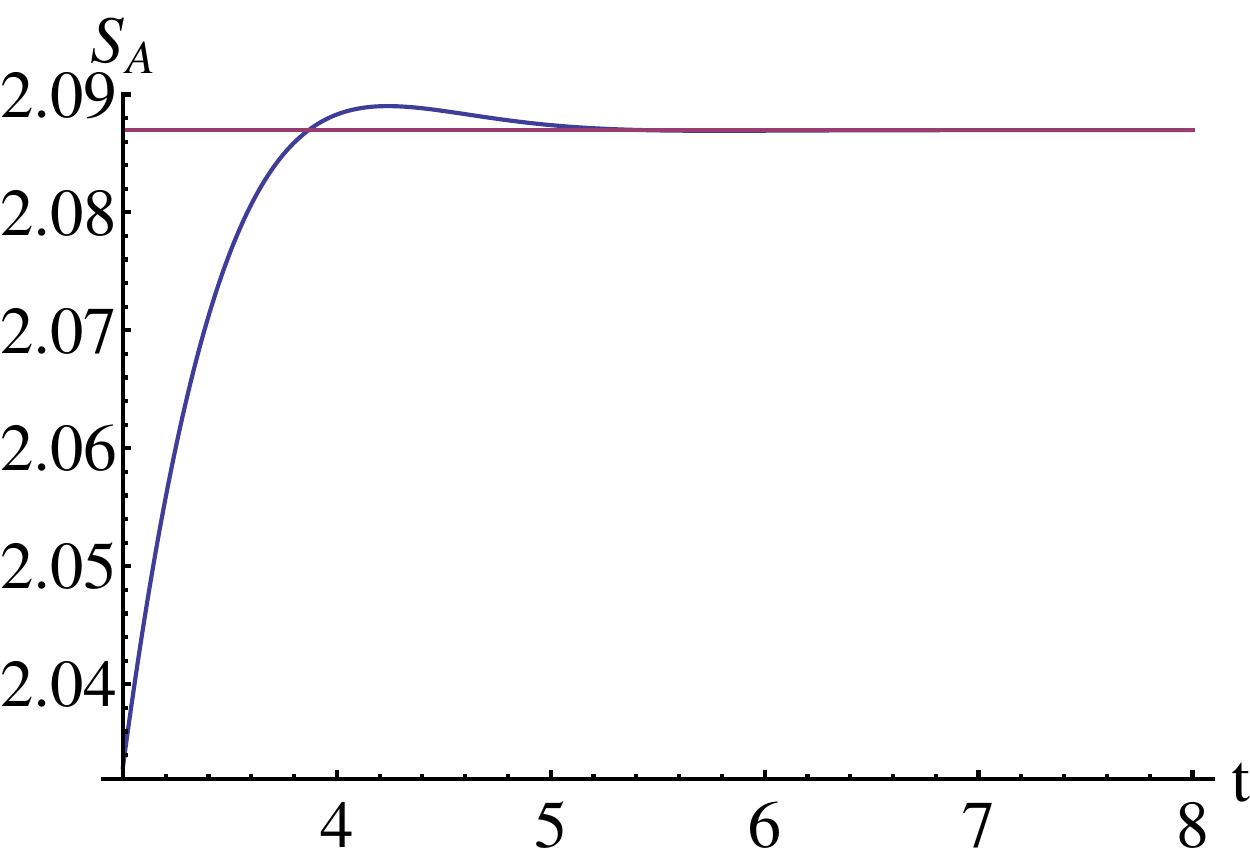}
  \caption{\small Non-monotonic behaviour, $\kappa_4=0.30$}
\end{subfigure}
\caption{\small Entanglement entropy growth of an interval(r=5) for different choice of $\kappa_4$ and $\kappa_2=1$.}
\label{gCC_EE}
\end{figure}

The most interesting aspect of Figure (\ref{gCC_EE}) is the non-monotonic growth of EE in the gCC state with `large' $\kappa_4$. As in case of CC state, to study the monotonic or non-monotonic behaviour of $S_A$, the more appropriate quantity is not $S_A$ but rather $\frac{\partial S_A}{\partial t}$, the expression also simplifies tremendously.
\begin{eqnarray}
\left<\frac{\partial S_A}{\partial t}\right>_{gCC}&=&\frac{1}{3}\int_{-\infty}^{\infty}\,dk\,(1-e^{ikr})\text{cosech}(2\kappa_2 k + 2\kappa_4 k^3)\sin(2kt)\nonumber\\
&=&\frac{1}{3}\int_{-\infty}^{\infty}\,dk\,(1-\cos(kr))\text{cosech}(2\kappa_2 k + 2\kappa_4 k^3)\sin(2kt)\
\label{dS_gCC}
\end{eqnarray}
Unfortunately, the above integral still cannot be done in closed form. The objective is to find finite positive real zeroes of the above expression as a function of time $t$. But, calculating zeroes of Fourier transforms, unless it can be done in closed form, is notoriously hard, the most famous example being the Riemann hypothesis.

The most interesting question that can be asked in Figure (\ref{gCC_EE}) is whether even a small infinitesimal $\kappa_4$, although not visible in the numerical plot, gives rise to the non-monotonic EE growth or whether the non-monotonic behaviour starts from a sharp finite value of $\kappa_4$. If it is the second case, then it is akin to a dynamical phase transition. In other words, the question is whether (\ref{dS_gCC}) has finite zeroes as a function of time even for an infinitesimal $\kappa_4$ or do the finite zeroes appear for $\kappa_4$ greater than a critical value.

\gap1

{\it{We found that the non-monotonic behaviour starts abruptly at a critical value of $\kappa_4=16 \kappa_2^3/27 \pi ^2$. In terms of the effective temperature and chemical potential in the stationary limit, $\beta = 4\kappa_2$ and $\mu_4=4\kappa_4$, the critical value is $\mu_{4c}=\beta^3/27 \pi ^2$.}}

\gap1

Althought the integral (\ref{dS_gCC}) cannot be done in closed form, we can take advantage of the fact that for our question we do not need to know the precise zeroes. Using contour integration, the integral is given by the sum of residues of the poles given by $2\kappa_2 k+2\kappa_4 k^3 =in\pi$ where $n\in\mathbb{Z}-\{0\}$. $n=0$ is not a pole of (\ref{dS_gCC}). The expressions of the poles(from \cite{Mandal:2015kxi})\footnote{The numerical values of the poles may get interchanged for specific values of the parameters but the result will always be the same set of roots. This arises from the particular method used for solving the cubic equation.} are
\begin{eqnarray}
k_1&=&\frac{-2\ 6^{2/3} \kappa_2+\sqrt[3]{6} \left(\sqrt{48 \kappa_2^3-81 \pi ^2 \kappa_4 n^2}+9 i \pi\sqrt{\kappa_4} n\right)^{2/3}}{6 \sqrt[3]{\sqrt{3} \sqrt{\kappa_4^3 \left(16 \kappa_2^3-27 \pi ^2\kappa_4 n^2\right)}+9 i \pi \kappa_4^2 n}}\\
k_2&=&\frac{4 \sqrt[3]{-6} \kappa_2+i \left(\sqrt{3}+i\right) \left(\sqrt{48 \kappa_2^3-81 \pi ^2 \kappa_4 n^2}+9 i\pi\sqrt{\kappa_4} n\right)^{2/3}}{2\ 6^{2/3} \sqrt[3]{\sqrt{3} \sqrt{\kappa_4^3 \left(16 \kappa_2^3-27\pi^2 \kappa_4 n^2\right)}+9 i \pi\kappa_4^2 n}}\\
k_3&=&-\frac{\sqrt[3]{-1} \left(2 \sqrt[3]{-6} \kappa_2+\left(\sqrt{48 \kappa_2^3-81 \pi ^2 \kappa_4 n^2}+9 i \pi \sqrt{\kappa_4} n\right)^{2/3}\right)}{6^{2/3} \sqrt{\kappa_4} \sqrt[3]{\sqrt{48 \kappa_2^3-81 \pi ^2\kappa_4 n^2}+9 i \pi  \sqrt{\kappa_4} n}}\
\label{poles}
\end{eqnarray}
Out of the three poles, only one is perturbative. In $\kappa_4\to 0$ series expansion, the other two start with $\mathcal{O}(\frac{1}{\sqrt{\kappa_4}})$. One of the three poles is always imaginary for arbitrary $n$ and arbitrary positive $\kappa_4$.

There are three important ingredients for the proof of the abrupt transition to non-monotonic behaviour:
\begin{enumerate}
\item All three $n^\text{th}$ poles become purely imaginary when $16 \kappa_2^3-27 \pi ^2\kappa_4 n^2$ is negative, or $\kappa_4$ is greater than $16 \kappa_2^3/27 \pi ^2 n^2$, we will call this the $n^{\text{th}}$ critical value $\kappa_{4c,n}$,
\begin{align}
\label{kappa4cn} \kappa_{4c,n}&= \frac{16 \kappa_2^3}{27 \pi ^2 n^2}\
\end{align}
Below this value, the residues of the $n^{\text{th}}$ poles are exponential decaying functions of time $t$, with no oscillatory factor. Obviously, ($n=\pm1$) critical\footnote{We will call this value just `critical value' without the `$n^{\text{th}}$' specification because, as shown below, this is the critical value of $\kappa_4$ where the dynamical phase transition happens.} value $\kappa_{4c}$ is larger than $\kappa_{4c,n}$ for $|n|>1$. With $\kappa$ scaled to $1$, $\kappa_{4c}$ is $16 \kappa_2^3/27 \pi ^2\sim 0.0600422$.
\item With $\kappa_4$ less than $(n=\pm1)$ critical value, the sum of the residues of ($n=\pm 1$) poles is larger than the sum of the residues of all the other ($|n|>1$) poles. Hence, the behaviour of the first poles of $n=\pm1$ dictate the behaviour of the integral (\ref{dS_gCC}) when $\kappa_4<16 \kappa_2^3/27 \pi ^2$.
\item Above this critical value, for each $n$, two of the poles have real parts while one of them, say $k_1$, is imaginary. The poles are
\begin{gather}
\label{poleR} k_1= -2i\,\text{sgn}(n)\,b, \qquad k_2=a+i\,\text{sgn}(n)\,b, \qquad k_3=-a+i\,\text{sgn}(n)\,b\\
a=\frac{B^{2/3}-2 \sqrt[3]{6} \kappa _2}{2\ 2^{2/3} \sqrt[6]{3} \sqrt[3]{B} \sqrt{\kappa _4}}, \qquad b= \frac{B^{2/3}+2 \sqrt[3]{6} \kappa_2}{2\ 6^{2/3} \sqrt[3]{B} \sqrt{\kappa_4}}\nonumber\\
B=\sqrt{81 \pi ^2 \kappa_4 n^2-48 \kappa_2^3}+9 \pi |n|\sqrt{\kappa_4}\nonumber\
\end{gather}
where we have to take the real roots of the radicals. $k_1$'s have the largest imaginary parts and the exponential decay of their residues as a function of time are faster while the other poles $k_2$ and $k_3$ have comparatively large magnitudes and ocsillations.\footnote{This competition between poles of each $n$ might be important, if we have turned on $W_6$ chemical potential instead of $W_4$, in which case there will be five poles, or $W_8$ in which case there will be seven poles and so on.} In the total integral, the contributions of the imaginary poles $k_1$'s cannot compete with the contributions of the oscillating poles.  Lastly, it would be a very special arrangment if all ocsillating terms conspire to give a non-oscillatory sum. Hence, the total integral is oscillatory as a function of time and the EE growth is non-monotonic.
\end{enumerate}
For future reference, we also note that the expansion of the real part `$a$' in (\ref{poleR}) around the $n^\text{th}$ critical value $\kappa_{4c,n}$ is
\begin{multline}
 a=\frac{\sqrt[3]{\pi } \sqrt[3]{|n|} \sqrt{\kappa _4-\kappa _{4 c,n}}}{2^{2/3} \sqrt{3} \kappa _{4 c,n}^{5/6}}-\frac{35 \left(\sqrt[3]{\pi } \sqrt[3]{|n|}\right) (\kappa_4-\kappa _{4 c,n})^{3/2}}{54 \left(2^{2/3} \sqrt{3} \kappa _{4 c,n}^{11/6}\right)}\\
 +\frac{1001 \sqrt[3]{\pi } \sqrt[3]{|n|} (\kappa_4-\kappa _{4 c,n})^{5/2}}{1944\ 2^{2/3} \sqrt{3} \kappa _{4 c,n}^{17/6}}+\mathcal{O}(\kappa_4-\kappa_{4c,n})^{7/2}
 \label{aexp}
\end{multline}

For all our calculations below, we have scaled $\kappa_2$ to be 1. The first point is clear from figure (\ref{RePol}). The real parts of ($n=\pm1$) poles vanish at $\kappa_4\sim 0.060$, which is the critical value found above. The critical value of $(n=\pm2$) poles is $\kappa_4\sim 0.015$.

\begin{figure}[H]
\centerline{\includegraphics[scale=.5]{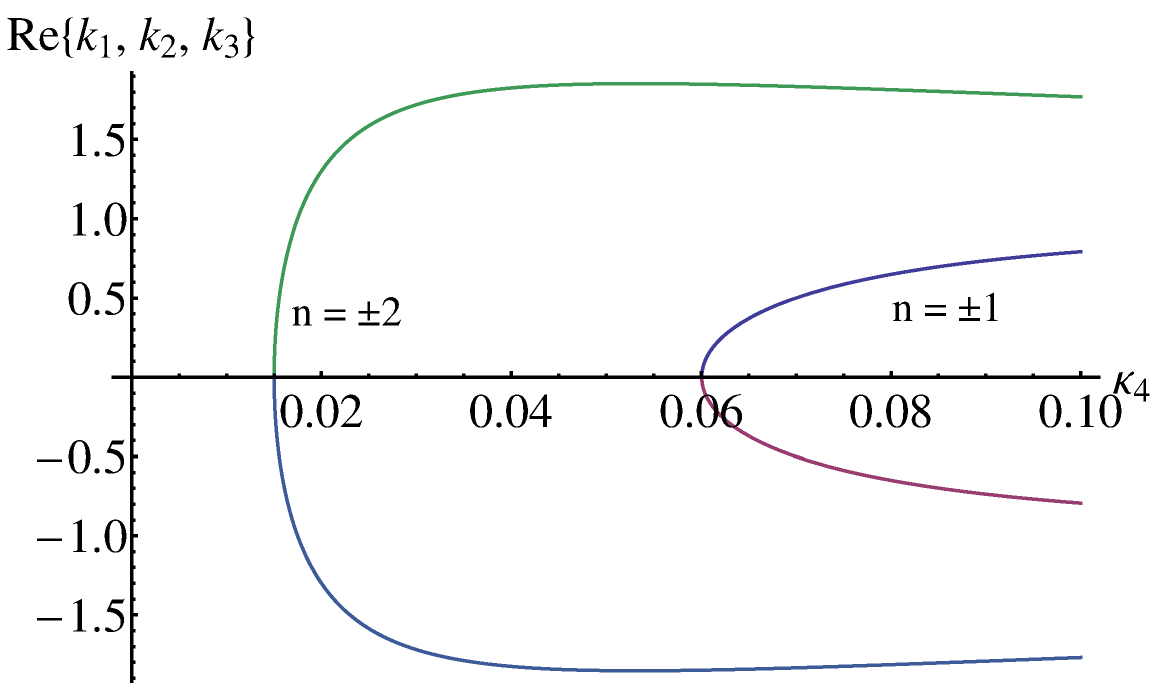}}
\caption{Real parts of poles of $n\in\{\pm1, \pm2\}$ as a function of $\kappa_4$ with $\kappa_2$ scaled to $1$.}
\label{RePol}
\end{figure}

Below the critical value, we will show that the total contributions from $n=\pm 1$ poles is larger than the sum of all residues of $|n|>1$ poles. We will concentrate on the late time period, $t>r/2$. For $e^{i2kt}$ of $\sin(2kt)$ factor in (\ref{dS_gCC}), the contour is closed upward encircling the upper half plane, and for $e^{-i2kt}$, the contour is closed downward encircling the lower half plane. From the expansion of $\csch\left(2 \kappa _4 \left(k-k_1\right)
\left(k-k_2\right) \left(k-k_3\right)+i \pi n\right)$ around $k_1$, the contribution from $k_1$ poles for arbitrary $n$ are the real parts of 
\begin{align}
P_n(k_1)&=2\pi iR_1(k_1)=\frac{(-1)^n}{6\kappa_4(k_1-k_2)(k_1-k_3)}\,\left(e^{i2k_1t}-\frac{e^{ik_1(r+2t)}+e^{ik_1(-r+2t)}}{2}\right) \; &\text{if} \;\text{Im}[k_1]>0\nonumber\\
\label{I1}\\
Q_n(k_1)&=-2\pi iR_2(k_1)=\frac{(-1)^n}{6\kappa_4(k_1-k_2)(k_1-k_3)}\,\left(e^{-i2k_1t}-\frac{e^{ik_1(r-2t)}+e^{-ik_1(r+2t)}}{2}\right)\; &\text{if} \;\text{Im}[k_1]<0\nonumber\\
\label{I2}\
\end{align}
where $R_1$ and $R_2$ denote the residues. Similarly, cyclic replacements of $k_1$ with $k_2$ and $k_3$ give the contributions of $k_2$ and $k_3$ poles. For the poles in the lower half of the complex plane, since the contour is anticlockwise, $Q_n$ have an extra minus sign in the residue.
We will call the contributions to the integral form $n=\pm1$ poles as $I_0(t)$ and the contributions of the $|n|>1$ poles as $I_1(t)$. The other parameters ($\kappa_4$, $r$ and $\kappa$ which is already scaled to 1) are suppressed.

As a first visual evidence, Figure (\ref{ccritNnR}) is the comparison of numerical integration of (\ref{dS_gCC}) and $I_0(t)$. It is evident that the residues of ($n=\pm1$) poles dominate the contour integration. We have chosen $\kappa_4=0.0600420$ which is very close to the critical value. As mentioned above, with this choice, all the poles except the $n=\pm 1$ poles give ocsillating residues as a function of time. Although it is not very conspicuous, it is also evident from the graph that $I_1(t)$ is oscillating around $I_0(t)$, the value of the numerical integration is above the $I_0(t)$ curve in some regions and below in other regions of time $t$.
\begin{figure}[H]
\centerline{\includegraphics[scale=.5]{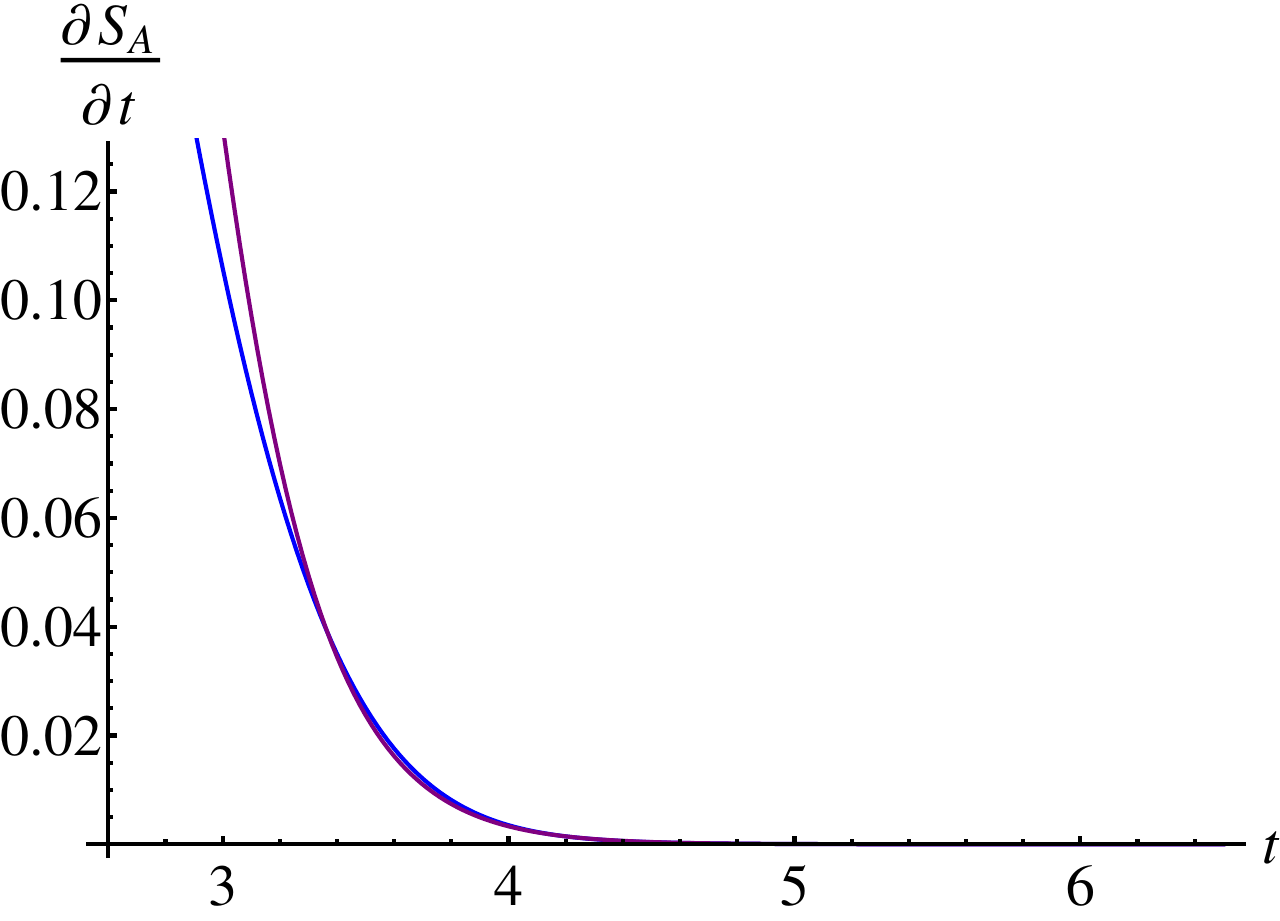}}
\caption{Comparison of numerical integration of $\left<\partial{S_A}/\partial t\right>_{gCC}$ (blue curve) and $I_0(t)$ (purple curve) as a function of time $t$. The parameters are $\kappa_4=0.0600420$, $r=5$.}
\label{ccritNnR}
\end{figure}
The numerical integration is unreliable in the long time limit. So, to complete our argument, we will calculate an upper bound of $I_1(t)$ and compare it with $I_0(t)$ for a specific time $t$. The choice of the parameters are
\begin{eqnarray}
\kappa=1, \; \kappa_4=0.0600420, \; r=5,\; t=4r=20,\
\label{para1}
\end{eqnarray}
With these parameters, the $n=1$ and $n=-1$ poles are
\begin{align}
 k_1&= 2.3538234 i & k_2&=2.3585719 i& k_3&=-4.7123954 i &;\; n&=1\\
 k_1&= 4.7123954 i& k_2&=-2.3538234 i& k_3&=-2.3585719 i &;\; n&=-1\
\end{align}
and $I_0(t)$ is given by
\begin{eqnarray}
 I_0(t)|_{t=20}&=&P(k_1)|_{n=1}+P(k_2)|_{n=1}+Q(k_3)|_{n=1}+P(k_1)|_{n=-1}+Q(k_2)|_{n=-1}+Q(k_3)|_{n=-1}\nonumber\\
 &=&6.646589\times10^{-35}\
 \label{Inpm1}
\end{eqnarray}
We can show that $I_1(t)|_{t=20}$ is less than $I_0(t)|_{t=20}$. The first few poles are
\begin{align*}
k_1&=5.5495551 i& k_2&=2.5383386 - 2.7747775 i & k_3&=-2.5383386 - 2.7747775 i  &;\; n&=-3\\
k_1&=5.1737935 i, & k_2&=1.8496206 - 2.5868967i & k_3&= -1.8496206 - 2.5868967 i &;\; n&=-2\\
k_1&=-5.1737935 i & k_2&= -1.8496206 + 2.5868967 i & k_3&=1.8496206 + 2.5868967 i &;\; n&=2\\
k_1&= -5.54955505 i & k_2&= -2.5383386 + 2.7747775 i & k_3&=2.5383386 +2.7747775 i &;\; n&=3\
\end{align*}
The residues of these ($|n|>1$) poles cannot be summed up into a closed form, as that would amount to doing the integral in closed form. We are interested in an upper bound. The residues of two of the three poles of every ($|n|>1$) have an oscillation factor. As we saw, even each residue has a separate 3-6 real oscillating terms as a function of time. So, we can represent the sum of the modulus (absolute value of the amplitude) of the oscillating terms of the three residues for each $n$, by a bigger function which has the analytic sum from $|n|>1$ to infinity. And if the sum is less than $I_0(t)$, then $I_0(t)$ dominates the contribution from all the other poles.\footnote{A simplified example of our strategy is the comparison between say $X$ and $a\sin(x)+b\cos(y)$ where $\{X,a,b,x,y\}\in\mathcal{R}$, while $A>|a|$ and $B>|b|$ and $\{A,B\}\in\mathcal{R}^+$, then $A+B>|a|+|b|>a\sin(x)+b\cos(y)$ and if $X>A+B$ then $X>a\sin(x)+b\cos(y)$.}
\begin{figure}[H]
\centerline{\includegraphics[scale=.6]{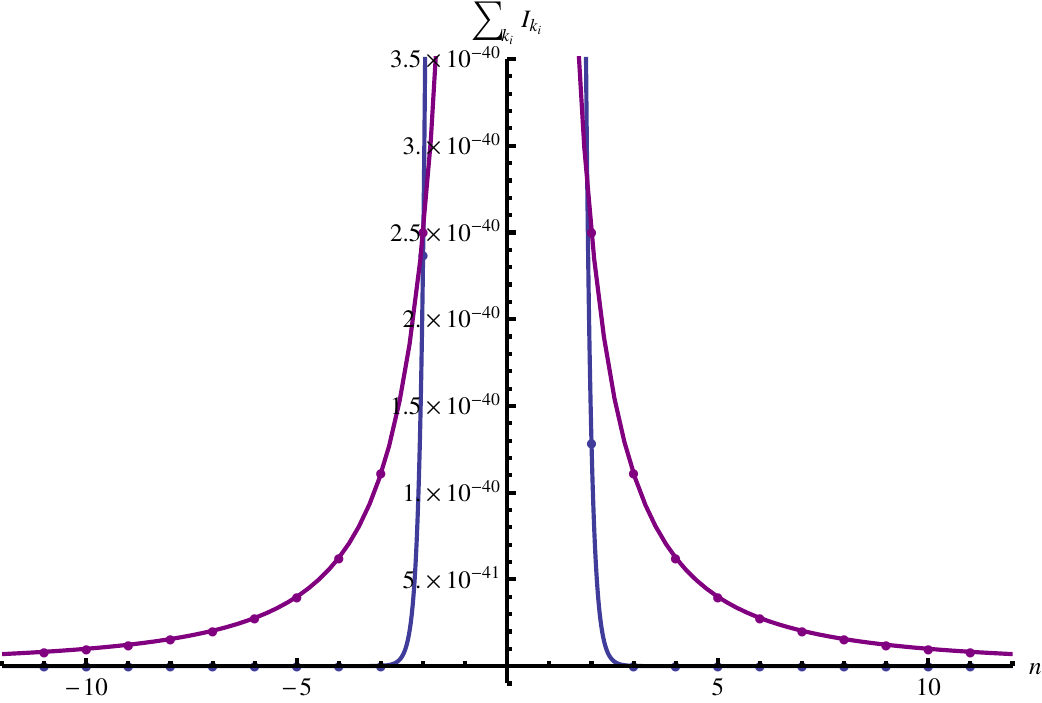}}
\caption{Comparison of sum of modulus of residues of ($|n|>1$) poles with the approximating function $f(n)=10^{-39}/n^2$. The dots are the discrete $n$ values of the corresponding functions.}
\label{ResComp}
\end{figure}
Figure (\ref{ResComp}) are the plots of the sum of the moduli separately for the oscillating terms of the three residues as a function of $n$ and the approximating function $f(n)=10^{-39}/n^2$. Now, we have
\begin{eqnarray}
 \sum_{-\infty}^{n=-2}\frac{10^{-39}}{n^2}+ \sum_{n=2}^{\infty}\frac{10^{-39}}{n^2} =1.289868\times10^{-39}\
 \label{rescompsum}
\end{eqnarray}
This is much less than $I_0(t)|_{t=20}$ in (\ref{Inpm1}) and is of the order of $10^{-5}$ of $I_0(t)|_{t=20}$. So, the non-oscillating $I_0(t)$ dominates $I_1(t)$, the contribution from the other poles. Hence, below $\kappa_4=16 \kappa_2^3/27 \pi ^2$, the EE growth is monotonic.

Visually from figure (\ref{ccritNnR}), $t=3.7$ is a time-slice where the difference between $I_0(t)$ and the numerical integration has a local maxima. At this time slice, repeating the above exercise, $I_0(t)|_{t=3.7}=0.109727$ and repeating the same exercise of estimating the upper bound of $I_1(t)|_{t=3.7}$ with the same parameters as (\ref{para1}) except the change in $t$, we get a good upper bound to be $0.0064493$ which is less than $I_0(t)|_{t=3.7}$ and is of the order of $60\%$ of $I_0(t)|_{t=3.7}$. So, the approximation of the full integral by $I_0(t)$ gets better with increasing time. In the long time limit, we can effectively take the only time-dependence to be the time-dependence of $I_0(t)$. {\it{It is worth mentioning here that even $(n=\pm1)$ pole calculations take into account $\kappa_4$ non-perturbatively, since two of the poles of each $n$ are non-perturbative in $\kappa_4$.}}
 
As listed above as one of the main points, above the critcal value, each $n$ has an imaginary pole but the other two poles have real parts and also have larger magnitudes so the total residue of the three poles of each $n$ is oscillatory. It would also be a very special arrangement if all the oscillatory contributions of each $n$ conspire to give a non-oscillatory $\partial S_A/\partial t$. Hence, we conclude that the EE growth is non-monotonic above the critical value.

Near the critical point $(\kappa_4-\kappa_{4c})\to 0^+$, we can try to estimate an upper bound of the time upto which the EE growth is monotonic. The upper bound is half of the longest time period. Using the leading term in expansion of `$a$' from (\ref{aexp}) and the expressions of the residues (\ref{I1}) and (\ref{I2}), the lowest frequency($|n|=1$) gives the upper bound as
\begin{align}
\frac{\sqrt[3]{\pi } \sqrt{\kappa _4-\kappa _{4 c}}}{2^{2/3} \sqrt{3} \kappa _{4 c}^{5/6}}(2t-r)=\pi\quad\Rightarrow\quad t= \frac{(2\pi)^{2/3} \sqrt{3} \kappa _{4 c}^{5/6}}{2\sqrt{\kappa _4-\kappa _{4 c}}}+\frac{r}{2}\;\sim\; \frac{2.95\,\kappa _{4 c}^{5/6}}{\sqrt{\kappa _4-\kappa _{4 c}}}\     
\end{align}
where finite `$r$' can be neglected in the limit $(\kappa_4-\kappa_{4c})\to 0^+$.

The critical value in terms the effective inverse temperature $\beta=4\kappa_2$ and chemical potential $\mu_4=4\kappa_4$ in the stationary limit is
\begin{eqnarray}
\mu_4=\frac{\beta^3}{27 \pi ^2}\
\label{mucrit}
\end{eqnarray}
\begin{figure}[H]
\centerline{\includegraphics[scale=.4]{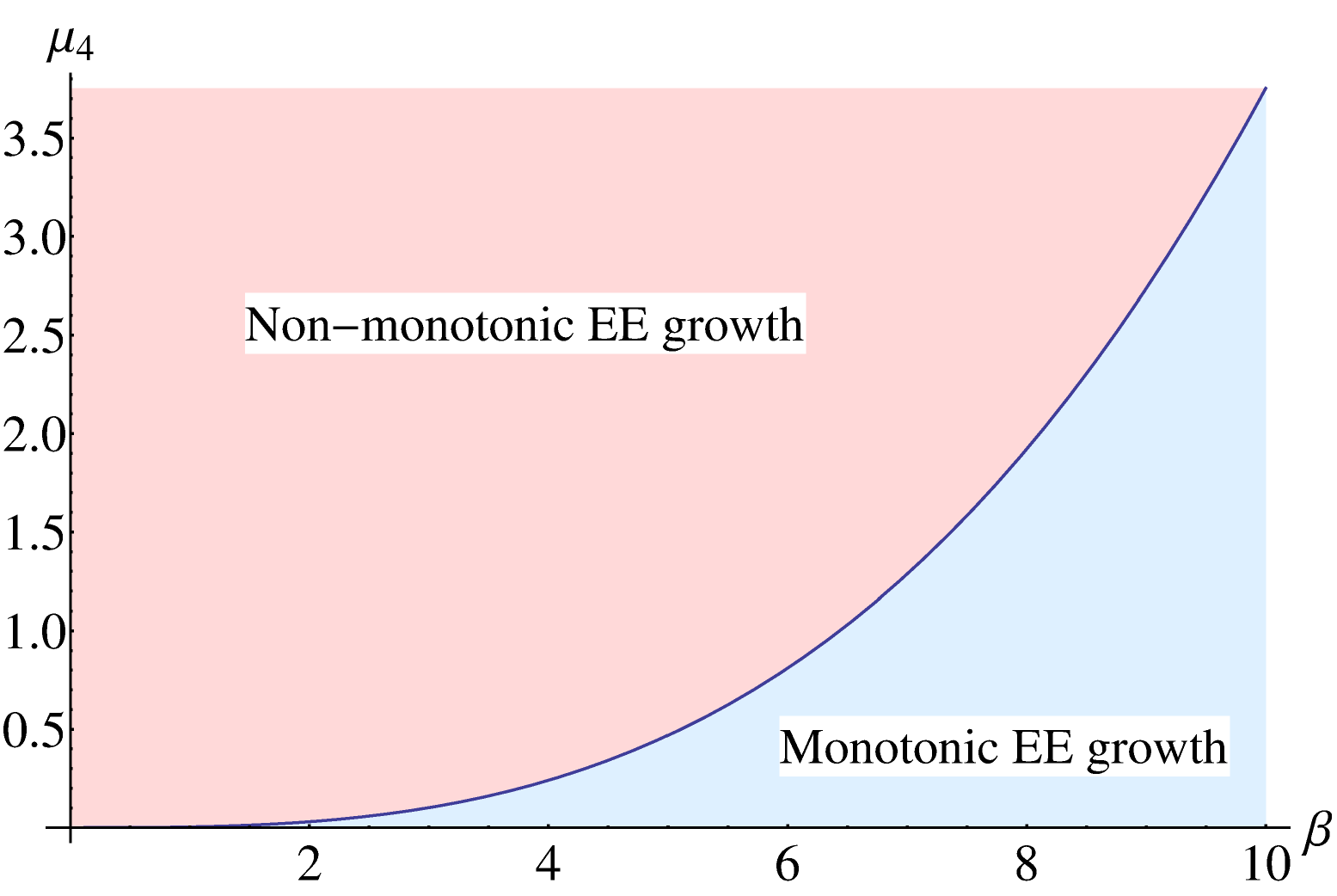}}
\caption{The critical curve $\mu_4=\beta^3/27 \pi ^2$ in terms of the effective inverse temperature $\beta$ and chemical potential $\mu_4$ in the stationary limit and the phase diagram.}
\label{pd}
\end{figure}
For the early times $t<r/2$, in the residue calculations (\ref{I1}) and (\ref{I2}), we have to replace the sign of the exponents with $r-2t$ so the magnitudes of the exponentials decreases as time increases. Upto the critical value of $\kappa_4$, the EE growth is always monotonic for this time period.

\subsection{Turning on other charges}
We could also calculate the EE growth of gCC states with other charges of the fermionic theory corresponding to bosonic bilinear $W_{2n}=\sum |k|^{2n-1}d^\dagger_kd_k$ where $n=3,4,5,...$. Repeating the exercise of quenching tuned squeezed states of scalar field theory in \cite{Mandal:2015kxi}, the propagator with these charges are simply given by
\begin{multline}
\langle\phi(0,t)\phi(r,t)\rangle=\int\frac{dk}{4\pi}\frac{e^{ikr}}{k} \left(\coth \left(2 \kappa_2 k +\sum_{n=2}^{\infty}\kappa_{2n}  k^{2n-1}\right)\right.\\
\left.-\cos (2 k t) \csch\left(2 \kappa_2 k +\sum_{n=2}^{\infty}\kappa_{2n}  k^{2n-1}\right)-1\right)
\end{multline}
Substituting this propagator in the general formula (\ref{RE}) and (\ref{EE}) give the R\'enyi entropy  and entanglement entropy. The first derivative of EE w.r.t. time is
\begin{eqnarray}
\left<\frac{\partial S_A}{\partial t}\right>_{gCC}=\frac{1}{3}\int_{-\infty}^{\infty}\,dk\,(1-\cos(kr))\text{cosech}\left(2\kappa k + \sum_{n=2}^{\infty}\kappa_{2n}  k^{2n-1}\right)\sin(2kt)\
\label{dS_gCCn} 
\end{eqnarray}
We believe the dynamics will be much richer with these other charges, with much more complex phase diagrams which can be in a $n-1$ dimensional space.
But the general poles analysis cannot be done in these cases because the poles will be given by quintic and higher order equations.
Considering gCC states with $W_4$ and $W_6$ charges, the numerical plots of EE growth looks the same as (\ref{gCC_EE}) where by trial and error method, some parameter subspace gives monotonic growth and some subspaces do not give monotonic growth. Considering $n=\pm1$, the poles are given by $2\kappa_2 k+2\kappa_4 k^3 + 2\kappa_6 k^5=i\pi$. For $\kappa_2=1$ and $\kappa_4=0.06$, numerically we find two interesting parameter subspaces of $\kappa_6$. The first one is when all the poles become imaginary when $\kappa_4$ is decreased.
\begin{align}
 &k_1 =2.0887597\,\text{sgn}(n)\,i, \quad  k_2=2.9527785\,\text{sgn}(n)\,i, \quad k_3=-6.5425830 \,\text{sgn}(n)\,i,\nonumber\\
 &k_4= -6.6158300 \,\text{sgn}(n)\,i, \quad k_5=8.1168748 \,\text{sgn}(n)\,i      \qquad\text{for}\qquad\kappa_6=0.0007249\\
 \nonumber\\
 &k_1=-0.0076887-6.5788763 \,\text{sgn}(n)\,i, \quad k_2=2.0887456 \,\text{sgn}(n)\,i, \quad k_3=2.9528549 \,\text{sgn}(n)\,i,\nonumber\\
 &k_4=8.1161520 \,\text{sgn}(n)\,i, \quad k_5=0.0076887-6.5788763 \,\text{sgn}(n)\,i     \qquad\text{for}\qquad\kappa_6=0.0007250\
\end{align}
This looks like the same transition if $n=\pm1$ dominates, but the poles with real parts have large imaginary part also, so they would be highly damped.
The  other case is
\begin{align}
 &k_1 =-0.8215058 + 1.9681831\,\text{sgn}(n)\,i, \quad  k_2=-5.2389645\,\text{sgn}(n)\,i, \quad k_3=-5.2472000 \,\text{sgn}(n)\,i,\nonumber\\
 &k_4= 6.5497983\,\text{sgn}(n)\,i, \quad k_5=0.8215058 + 1.9681831 \,\text{sgn}(n)\,i    \qquad\text{for}\qquad\kappa_6=0.0019179\\
 &k_1=-0.8215060 + 1.9681836 \,\text{sgn}(n)\,i, \quad k_2=-0.0040372 - 5.2430724\,\text{sgn}(n)\,i, \nonumber\\
 &k_3=6.5497775 \,\text{sgn}(n)\,i, \quad k_4=0.0040372 - 5.2430724\,\text{sgn}(n)\,i, \nonumber\\
 &k_5=0.8215060 + 1.9681836\,\text{sgn}(n)\,i    \qquad\text{for}\qquad\kappa_6=0.0019180\
\end{align}
for the smaller $\kappa_4$, although two of the poles have real parts, they have to compete with the three imaginary poles. So, this could also be phase transition.

\section{Non-monotonic growth of entanglement from long range correlation at early times}
\label{secEE_gCC_arg}
In this section, we provide evidences to support our claim that the non-monotonic growth of entanglement entropy is due to long(er) range correlation at early times. First, we show that the non-monotonic growth is a boundary effect - no dependence on the size of the subsystem. Figure \ref{fig:maxSAminusSAtherm} is the plot of difference between the maximum $S_A(t)$ and the long time therm $S_A$ as a function of the subsystem size $r$. The plot is for different fixed values of $\kappa_4$ taking $\kappa_2=1$. We can see that the plots do not depend on the subsystem size (for relatively large systems $r\sim\beta=4\kappa_2$). In the rest of this section, we will work with $\kappa_2=1$. We also work with values of $\kappa_4$ near the transition value $\kappa_{4c}=0.06$ otherwise the other oscillatory terms that we have mentioned in the above section comes into play and makes the dynamics more complicated.

Next we show that the long range correlation increases while the short range correlation decreases. Figure \ref{fig:longer_correlation} are plots of the difference of $\langle J(0,0)\bar{J}(r,0)\rangle$ with non-zero $\kappa_4$ and without $\kappa_4$ as a function of the separation $r$ at time $t=0$. $J$ and $\bar{J}$ are the U(1) currents defined in (\ref{JbJdef}).
\begin{equation}
\delta(\langle J(0,0)J(r,0)\rangle)=\langle J(0,0)\bar{J}(r,0)\rangle_{\kappa_4}-\langle J(0,0)\bar{J}(r,0)\rangle_{\kappa_4=0}
\end{equation}
We can see that the long range correlation (away from $r=0$) increases with increasing $\kappa_4$ while short range correlation (near $r=0$) decreases. The changes in the quantum correlation pattern of the excitations at the boundaries of the interval gives rise to the non-monotonic growth. Note that the position $r$ of the maximum value of the plots in Figure \ref{fig:longer_correlation} increases as we increase $\kappa_4$. This means that the early time quantum correlation is spread further with increasing $\kappa_4$.

Lastly we show that the maximum of $S_A$ reaches earlier as we increase $\kappa_4$. This is visible from Figure \ref{fig:t_at_maxSA} which is a plot of $t_{max}$ (time at which $S_A(t)$) is maximum) as a function of varying $\kappa_4$. $t_{max}$ decreases with increasing $\kappa_4$. This observation agrees with the above observation that the early time quantum correlation is spread further with increasing $\kappa_4$. We can see that the plot rises again for $\kappa_4\geq 0.24$. This value of $\kappa_4$ is much larger than the critical value $\kappa_{4c}=0.06$ so the other oscillatory terms are coming to play.

\begin{figure}
  \centering
  \begin{subfigure}{.5\textwidth}
  \includegraphics[width=.9\linewidth]{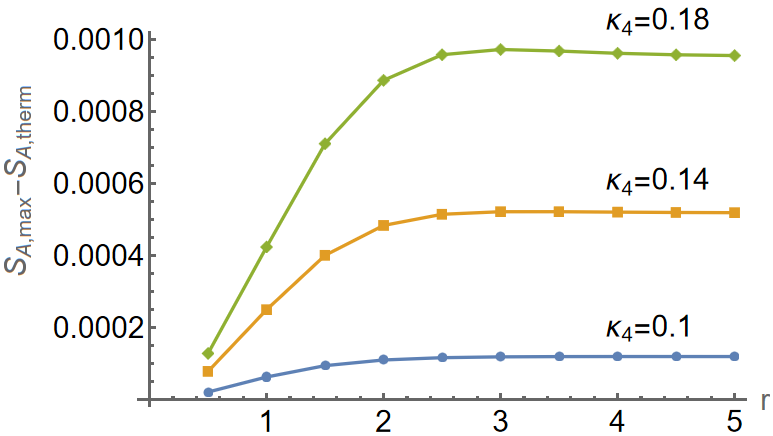}
  \caption{\small Max($S_A$) - $S_A(t \to \infty)$}
  \label{fig:maxSAminusSAtherm}
\end{subfigure}%
\begin{subfigure}{.5\textwidth}
  \centering
  \includegraphics[width=.9\linewidth]{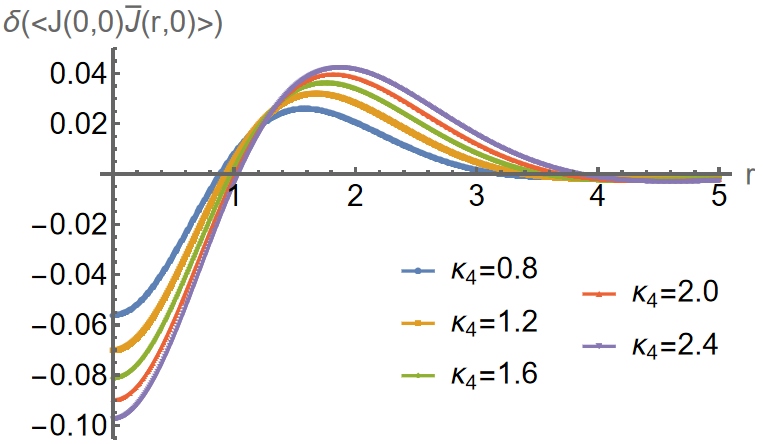}
  \caption{\small $\langle J(0,0)\bar{J}(r,0)\rangle_{\kappa_4}-\langle J(0,0)\bar{J}(r,0)\rangle_{\kappa_4=0}$}
  \label{fig:longer_correlation}
\end{subfigure}
\caption{\small (a) Plot of Max($S_A$) - $S_A(t \to \infty)$ as a function of the subsystem size $r$. This shows that the non-monotonic growth is a boundary effect. (b) Plot of $\delta(\langle J(0,0)J(r,0)\rangle)=\langle J(0,0)\bar{J}(r,0)\rangle_{\kappa_4}-\langle J(0,0)\bar{J}(r,0)\rangle_{\kappa_4=0}$ as a function of separation distance $r$. Larger value of $\kappa_4$ leads to spreading of quantum correlation at early times.}
\label{fig:longerangecorrelation1}
\end{figure}

\begin{figure}
  \centering
  \includegraphics[width=.6\linewidth]{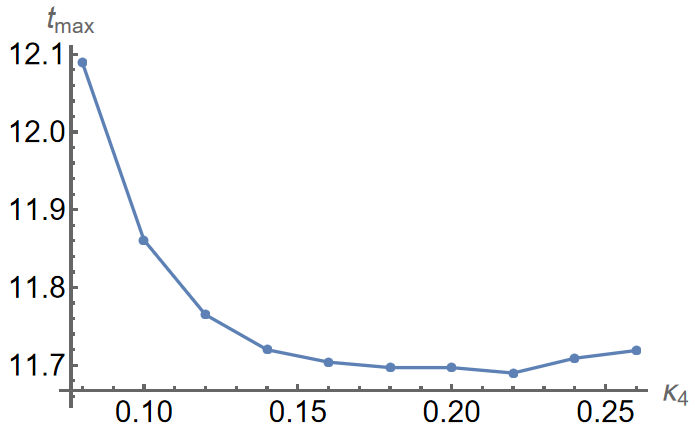}
  \caption{\small $t_{max}$ (time at which $S_A(t)$ is maximum) as a function of $\kappa_4$. $\kappa_2=1$ is fixed.}
\label{fig:t_at_maxSA}
\end{figure}

In summary, these evidences support our claim that the non-monotonic growth of entanglement entropy is due to increase in long range correlation and decrease in short range correlation at early times. We made attempts to modify the simple quasiparticle picture to take this into account and try to reproduce, at least qualitatively, the non-monotonic growth of entanglement entropy. But, so far, all such attempts failed to reproduce the non-monotonic growth. We believe this is due to intricate interplay of the increase in longer range correlation and decrease in short range correlation. The gradient of the fall-off of the correlation needs to be taken care of in detail. On the other hand, we believe that the more familiar bilinear $W_{2n}$ charges of the fermionic theory could also give rise to non-monotonic growth of entanglement entropy because of the higher derivatives involved which would give rise to longer range correlation at early times.

\section{Conclusions}
\label{cond}
In this paper, we have examined free fermionic mass quench. We find that the ground state quench equilibrates but not to a thermal ensemble. Starting from specially prepared squeezed states, we get CC and gCC states with fermionic bilinear $W_{2n}$ charges. Calculation of correlators in CC and gCC states explicitly shows thermalization to thermal emsemble and GGE respectively.

For CC state, we calculate EE growth exactly. The EE growth is strictly monotonically increasing. For gCC state with a particular charge, we find non-monotonic growth of entanglement entropy and subsequent thermalization. This happens when the chemical potential is greater than a sharp critical value. We argue that the non-monotonic growth is due to to increase in long range correlation and decrease in short range correlation at early times. So, we expect that the more familiar bilinear $W_{2n}$ charges of free fermionic theory could also give rise to non-monotonic growth of entanglement entropy because of the higher derivatives involved which would give rise to longer range correlation at early times. It would be interesting to reproduce our result in large $c$ holographic CFTs and examine what it would mean for Black hole physics.

\section*{Acknowledgement}
We thank Gautam Mandal for many discussions on this work and for proofreading the draft. This work is possible because of scholarship grants from the Government of India. This work was also partly supported by Infosys Endowment for the study of the Quantum Structure of Space Time.

\appendix

\section{\label{conv}Conventions}
\begin{eqnarray}
&&\eta_{\mu\nu}= \begin{bmatrix}    1 & 0 \\ 0 & -1  \end{bmatrix}, \quad \partial_{\mu} = (\partial_t, \partial_x), \quad \gamma^\mu\partial_\mu=\gamma^0\partial_t-\gamma^1\partial_x,\nonumber\\
&&w=t-x, \quad \bar{w}=t+x, \quad \partial=\frac{\partial}{\partial w}=\frac{1}{2}\left(\frac{\partial}{\partial t}-\frac{\partial}{\partial x}\right), \quad \bar{\partial}=\frac{\partial}{\partial \bar{w}}=\frac{1}{2}\left(\frac{\partial}{\partial t}+\frac{\partial}{\partial x}\right)\nonumber\\
&&\gamma^0_d=\begin{bmatrix}    1 & 0 \\ 0 & -1  \end{bmatrix}, \quad\gamma^1_d=\begin{bmatrix}    0 & 1 \\ -1 & 0  \end{bmatrix}, \quad\text{in Dirac basis}.\nonumber\\
&&S=\frac{1}{\sqrt{2}}\begin{bmatrix}    1 & -1 \\ 1 & 1  \end{bmatrix}, \gamma^0_c=S\gamma^0_dS^{-1}=\begin{bmatrix}    0 & 1 \\ 1 & 0  \end{bmatrix}, \quad \gamma^1_c=S\gamma^1_dS^{-1}=\begin{bmatrix}    0 & 1 \\ -1 & 0  \end{bmatrix},\quad\text{in chiral basis.}\nonumber\\ 
&&\left\{a_k,a^\dagger_{k'}\right\}=2\pi\,\delta(k-k'),\quad \left\{b_k,b^\dagger_{k'}\right\}=2\pi\,\delta(k-k'), \, \text{other anticommutators are zero.}\nonumber\\
&&\left\{a_n,a^\dagger_{n'}\right\}=\delta(n-n'),\quad \left\{b_n,b^\dagger_{n'}\right\}=\delta(n-n'), \, \text{other anticommutators are zero.}\nonumber\
\end{eqnarray}
We will use $k=\frac{2\pi n}{L}$ for continuum limit($L\to\infty$) and $n$ for quantization in a finite box of size L, where $n=n'+1/2$ and $n'\in\mathbb{Z}$.

\section{\label{spbas}Spinors and transformation to chiral basis:}
Taking constant mass $m$, we can easily find the boosted spinors, $u(k, m)$ and $v(k,m)$. For constant $m$, $\phi_{+p}(t)=e^{-i\omega t}$ and $\phi_{-m}(t)=e^{i\omega t}$. So from (\ref{spinor_antz}),
\begin{eqnarray}
 U(x,t)&=&\left[ \gamma^0\partial_t-\gamma^1\partial_x -im\right]e^{-i\omega t +ikx}\begin{bmatrix} 1 \\ 0  \end{bmatrix}=i \begin{bmatrix} -(\omega+m) \\ k  \end{bmatrix}e^{-ik\cdot x}\nonumber\\
 V(x,t)&=&\left[ \gamma^0\partial_t-\gamma^1\partial_x -im\right]\begin{bmatrix} 0 \\ 1  \end{bmatrix}e^{ik\cdot x}=i \begin{bmatrix} k \\ -(\omega+m)  \end{bmatrix}e^{ik\cdot x}\nonumber\
\end{eqnarray}
Hence, upto normalizations fixed by inner products, the boosted spinors are
\begin{eqnarray}
 u(k,m)=i \begin{bmatrix} -(\omega+m) \\ k  \end{bmatrix}, \quad v(k,m)=i \begin{bmatrix} k \\ -(\omega+m)  \end{bmatrix}\nonumber\
\end{eqnarray}
We have the adjoint spinors as,
\begin{eqnarray}
 \bar{u}(k,m)&=&u^\dagger(k,m)\gamma^0=-i \begin{bmatrix} -(\omega+m) & k  \end{bmatrix} \begin{bmatrix}    1 & 0 \\ 0 & -1  \end{bmatrix}=i\begin{bmatrix} -(\omega+m) & -k  \end{bmatrix}\nonumber\\
 \bar{v}(k,m)&=&v^\dagger(k,m)\gamma^0= -i \begin{bmatrix} k & -(\omega+m)  \end{bmatrix}\begin{bmatrix}    1 & 0 \\ 0 & -1  \end{bmatrix}=i\begin{bmatrix} k & (\omega+m)  \end{bmatrix}\nonumber\
\end{eqnarray}
Now borrowing Peskin \& Schroeder \cite{Peskin:1995ev} conventions of spinors, we want to fix the inner products $\bar{u}(k,m)u(k,m)=2m$ and $\bar{v}(k,m)v(k,m)=-2m$,
\begin{eqnarray}
 \bar{u}(k,m)u(k,m)&=&\begin{bmatrix} -(\omega+m) & -k  \end{bmatrix}\begin{bmatrix} -(\omega+m) \\ k  \end{bmatrix}=2m(\omega+m)\nonumber\\
 \bar{v}(k,m)v(k,m)&=&\begin{bmatrix} k & (\omega+m)  \end{bmatrix}\begin{bmatrix} k \\ -(\omega+m)  \end{bmatrix}=-2m(\omega+m)\nonumber\
\end{eqnarray}
So the normalized spinors are
\begin{eqnarray}
u(k,m)&=&\frac{1}{\sqrt{(\omega+m)}}\begin{bmatrix} (\omega+m) \\ -k  \end{bmatrix}, \quad v(k,m)=\frac{1}{\sqrt{(\omega+m)}} \begin{bmatrix} k \\ -(\omega+m)  \end{bmatrix}\nonumber\\
\bar{u}(k,m)&=&\frac{1}{\sqrt{(\omega+m)}}\begin{bmatrix} (\omega+m) & k  \end{bmatrix}, \quad \bar{v}(k,m)=\frac{1}{\sqrt{(\omega+m)}}\begin{bmatrix} k & (\omega+m)  \end{bmatrix}\nonumber\\
\label{norspinor}
\end{eqnarray}
The spinors with time-dependent mass $m(t)$ are obtained by just substituting $m(t)$ in the place of `$m$' only inside the matrices, which is clearly seen from (\ref{Uxt}) and (\ref{Vxt}). The normalization cannot be changed to time-dependent mass else the spinors won't be solutions of the corresponding Dirac equation.

The transformation to chiral basis is accomplished by using the transformation matrix $S=\frac{1}{\sqrt{2}}\begin{bmatrix} 1 & -1 \\ 1 & 1 \end{bmatrix}$. The mode expansion as in Peskin \& Schroeder is
\begin{eqnarray}
 \Psi(x,t)&=&\int \frac{dk}{2\pi}\frac{1}{\sqrt{2\omega}}\left[ a_k u(k,m) e^{-ik\cdot x}+b^\dagger_k v(k,m) e^{ik\cdot x}\right]\nonumber\\
 &=& \int \frac{dk}{2\pi}\frac{1}{\sqrt{2\omega}}\left[ a_k \frac{1}{\sqrt{(\omega+m)}}\begin{bmatrix} (\omega+m) \\ -k  \end{bmatrix} e^{-ik\cdot x}+b^\dagger_k \frac{1}{\sqrt{(\omega+m)}} \begin{bmatrix} k \\ -(\omega+m)  \end{bmatrix} e^{ik\cdot x}\right]\nonumber\\
 &\xrightarrow{m\to 0}&  \int \frac{dk}{2\pi}\frac{1}{\sqrt{2}}\left[ a_k\begin{bmatrix} 1 \\ -\text{sgn}(k)  \end{bmatrix} e^{-ik\cdot x}+b^\dagger_k \begin{bmatrix} \text{sgn}(k) \\ -1  \end{bmatrix} e^{ik\cdot x}\right]\nonumber\\
 &=& \int \frac{dk}{2\pi}\frac{1}{\sqrt{2}}\begin{bmatrix} a_k e^{-ik\cdot x}+\text{sgn}(k)b^\dagger_k e^{ik\cdot x} \\ -\text{sgn}(k) a_k e^{-ik\cdot x} -b^\dagger_k e^{ik\cdot x} \end{bmatrix}\nonumber\
\end{eqnarray}
In the chiral basis,
\begin{eqnarray}
 \Psi_c(x,t)&=&S\cdot\Psi(x,t)=\frac{1}{\sqrt{2}}\begin{bmatrix} 1 & -1 \\ 1 & 1 \end{bmatrix}\cdot \int \frac{dk}{2\pi}\frac{1}{\sqrt{2}}\begin{bmatrix} a_k e^{-ik\cdot x}+\text{sgn}(k)b^\dagger_k e^{ik\cdot x} \\ -\text{sgn}(k) a_k e^{-ik\cdot x} -b^\dagger_k e^{ik\cdot x} \end{bmatrix}\nonumber\\
 &=& \int_{-\infty}^{\infty} \frac{dk}{2\pi}\frac{1}{2}\begin{bmatrix} (1+\text{sgn}(k))(a_k e^{-ik\cdot x}+b^\dagger_k e^{ik\cdot x}) \\ (1-\text{sgn}(k))(a_k e^{-ik\cdot x}-b^\dagger_k e^{ik\cdot x}) \end{bmatrix}\
 \label{psic}
\end{eqnarray}
Writing this as $\psi(x,t)$ and $\bar{\psi}(x,t)$,
\begin{eqnarray}
\label{crep} \psi(x,t)&=&\int_0^{\infty} \frac{dk}{2\pi}(a_k e^{-ik\cdot x}+b^\dagger_k e^{ik\cdot x})\\
 \bar{\psi}(x,t)&=&\int_{-\infty}^0 \frac{dk}{2\pi}(a_k e^{-ik\cdot x}-b^\dagger_k e^{ik\cdot x})\
 \label{crepb}
\end{eqnarray}
\section{\label{bstate} Fermionic Boundary State}
The action (\ref{action}) with $m(t)=0$ in the chiral basis is
\begin{align*}
S&=-\int dx^2\,\left[i\bar{\Psi}\gamma^\mu\partial_\mu\Psi+\Psi\gamma^\mu\partial_\mu\bar{\Psi}\right]\\
&=-\frac{i}{2}\int dwd\bar{w} \left(\psi^\dagger \bar{\partial}\psi + \bar{\psi}^\dagger \partial \bar{\psi} + \psi \bar{\partial}\psi^\dagger + \bar{\psi} \partial \bar{\psi}^\dagger\right) \
\end{align*}

On varying the action and collecting terms, we get the following
\begin{equation}
\delta S = \int d^2x \left(\delta \psi^\dagger \bar{\partial}\psi + \delta\psi \bar{\partial}\psi^\dagger + \delta \bar{\psi}^\dagger \partial \bar{\psi} + \delta\bar{\psi} \partial \bar{\psi}^\dagger\right) 
\end{equation}
Given a boundary at $t=0$, it will also have certain boundary terms, which we want to be zero.
\begin{equation}
\left.\psi^\dagger\delta\psi + \psi\delta\psi^\dagger+\bar{\psi}^\dagger \delta \bar{\psi} + \bar{\psi} \delta \bar{\psi}^\dagger\right|_{t=0}=0
\end{equation}
We impose this as an operator equation on the boundary state $\left|B\right>$. The condition for a boundary state can be achieved via two identifications
\begin{gather}
\label{b1}\psi=i\bar{\psi}^\dagger,\quad\text{and}\quad \psi^\dagger=i\bar{\psi}\\
\label{b3}\psi=-\bar{\psi},\quad\text{and}\quad \psi^\dagger=\bar{\psi}^\dagger\
\end{gather}
Now we impose the boundary conditions at $t=0$ in terms of the mode expansions (\ref{crep}) and (\ref{crepb}):
\begin{enumerate}
\item The boundary condition of (\ref{b1}) gives $a_k\mp ia_{-k}^\dagger=0$ for $k>0$ and $b_k \mp ib_{-k}^\dagger=0$ for $k<0$. 
Similarly, the second condition is $a_k\pm ia_{-k}^\dagger=0$ for $k<0$ and $b_k\pm ib_{-k}^\dagger=0$ for $k>0$. Combining the separate conditions, we get $a_k\mp i\,\text{sgn}(k)a^\dagger_{-k}=0$ and $b_k\pm i\, \text{sgn}(k)b_{-k}^\dagger=0$. Hence, the boundary state corresponding to the first identification is
\begin{equation}
\left|N\right> = \exp\left(\sum\limits_k \ i\sgn(k) (a^\dagger_ka^\dagger_{-k}-b^\dagger_kb^\dagger_{-k})\right)\left|0\right>
\end{equation}
\item The boundary condition (\ref{b3}) is $a_k \mp\text{sgn}(k) b_{-k}^\dagger=0$ for $k>0$, $a_k\pm b_{-k}^\dagger=0$ for $k<0$ and $b_k \pm a_{-k}^\dagger=0$ for $k<0$ and $b_k\mp a_{-k}^\dagger=0$ for $k>0$. The boundary state for the first identification is
\begin{equation}
\left|D\right> = \exp\left(\sum\limits_k \sgn(k) a^\dagger_kb^\dagger_{-k}\right)\left|0\right>
\end{equation}
\end{enumerate}
From the action $S$, we can find the non-zero components of the energy-momentum tensor $T=T_{ww}$ and $\bar{T}=T_{\bar{w}\bar{w}}$, and the components of the $U(1)$ current are $J_w=J$ and $J_{\bar{w}}=\bar{J}$,
\begin{align}
&T=\frac{i}{2}\left(\psi^\dagger\partial\psi+\psi\partial\psi^\dagger\right) & \bar{T}&=\frac{i}{2}\left(\bar{\psi}^\dagger\bar{\partial}\bar{\psi}+\bar{\psi}\bar{\partial}\bar{\psi}^\dagger\right)\\
&J=\psi^\dagger\psi & \bar{J}&=\bar{\psi}^\dagger\bar{\psi}\
\label{JbJdef}
\end{align}
The boundary conditions (\ref{b1}) and (\ref{b3}) satisfy the condition
\begin{align}
 T(w)|_{t=0}&=\bar{T}(\bar{w})|_{t=0}, \qquad \text{on the cylinder.}\\
 \text{or},\quad \left(z^2T_{zz}(z)\right)|_{z(t=0)}&=\left(\bar{z}^2\bar{T}_{\bar{z}\bar{z}}(\bar{z})\right)|_{\bar{z}(t=0)}, \qquad \text{on the plane.}\
\end{align}
where $z=e^{2\pi(t-ix)/L}$ and $\bar{z}=e^{2\pi(t+ix)/L}$. Thus $\left|N\right>$ and $\left|D\right>$ are conformal invariant boundary states.
It is also worth noting that the boundary conditions also satisfy
\begin{align}
 J(w)|_{t=0}&=-\bar{J}(\bar{w})|_{t=0}, \qquad \text{on the cylinder.}\\
 \text{or},\quad \left(zJ_z(z)\right)|_{z(t=0)}&=\left(\bar{z}\bar{J}_{\bar{z}}(\bar{z})\right)|_{\bar{z}(t=0)}, \qquad \text{on the plane.}\
\end{align}
Considering the zero modes in the cylinder, it means that the above boundary states are not charged. With $Q=J_0+\bar{J}_0$,
\begin{eqnarray}
Q| N\rangle=0, \qquad Q| D\rangle=0\
\label{Q0}
\end{eqnarray}
Besides, specially for the state $|D\rangle$, $(J_0-\bar{J}_0)|D\rangle=0$. Hence
\begin{eqnarray}
J_0|D\rangle=0 \qquad \bar{J}_0|D\rangle=0\ 
\label{JbJ0}
\end{eqnarray}

\section{\label{bbs} Bosonised Boundary State}
Consider a Dirichlet boundary state $\varphi\left|D\right>=-\bar{\varphi}\left|D\right>$. Using the bosonised fermions :
\begin{align*}
\psi &= e^{-i\frac{\sqrt{\pi}}{4}\bar{P}} :e^{-i\sqrt{4\pi}\varphi(w)}: & \psi^\dagger &= e^{i\frac{\sqrt{\pi}}{4}\bar{P}} :e^{i\sqrt{4\pi}\varphi(w)}:\\
\bar{\psi} &= e^{-i\frac{\sqrt{\pi}}{4}P} :e^{i\sqrt{4\pi}\bar{\varphi}(\bar{w})}: & \bar{\psi}^\dagger &= e^{i\frac{\sqrt{\pi}}{4} P} :e^{-i\sqrt{4\pi}\bar{\varphi}(\bar{w})}:\
\end{align*}

To translate the boson Dirichlet condition into the fermionic one, we get
\begin{eqnarray}
\psi\left|D\right>&=& e^{-i\frac{\sqrt{\pi}}{4}\bar{P}} :e^{-i\sqrt{4\pi}\varphi}:|D\rangle\nonumber\\
&=&e^{-\frac{\pi}{2}[Q,P]} e^{-i\frac{\sqrt{\pi}}{4}\bar{P}} e^{-i\frac{\sqrt{\pi}}{4}P}:e^{-i\sqrt{4\pi}\varphi}:\left|D\right>\nonumber\\
&=&e^{-\frac{\pi}{2}[Q,P]} e^{-i\frac{\sqrt{\pi}}{4}\bar{P}} e^{-i\frac{\sqrt{\pi}}{4}P}:e^{i\sqrt{4\pi}\bar{\varphi}}:\left|D\right>\nonumber\\
&=& e^{\frac{\pi}{2}([Q,P]-[\bar{P},\bar{Q}])}\ \bar{\psi}\left|D\right>\nonumber\\
&=& e^{-i\pi}\ \bar{\psi} \left|D\right>= -\bar{\psi}\left|D\right>\nonumber\
\end{eqnarray}
where we have used the relation $e^x\,e^y=e^{y+[x,y]+...}\,e^x$ and $[Q,P]=[\bar{Q},\bar{P}]=i$. We have also used (\ref{JbJ0}) which gives $P|D\rangle=J_0|D\rangle=0$ and $\bar{P}|D\rangle=\bar{J}_0)|D\rangle=0$. 

Similarly, we can show that $(\psi^\dagger-\bar{\psi}^\dagger)\left|D\right>$, $(\psi-i\bar{\psi}^\dagger)\left|N\right>$ and $(\psi^\dagger-i\bar{\psi})\left|N\right>$ vanish, where $\left|N\right>$ is defined by $(\varphi - \bar{\varphi})\left|N\right> =0$ which is the Neumann boundary condition for scalar fields.

\section{\label{bch}Baker-Campbell-Hausdorff(BCH) formula}
Although we are interested in the `out' massless oscillators, the BCH formula is valid for both massive and massless oscillators. So, we will suppress the `in' or `out' identification of the oscillators. Starting from
\begin{eqnarray}
 |\Psi\rangle=\exp\left(\sum_k \text{sgn}(k)\gamma(k)a^\dagger_{k}b^\dagger_{-k}\right)|0\rangle\
\end{eqnarray}

we wish to obtain an expression of the form
\begin{equation}
\label{BCHeqn}\left|\psi\right> = \exp\left(-\sum\limits_k \kappa(k) (a^\dagger_k a_k + b^\dagger_k b_k)\right)\exp\left(\sum_k \sgn(k) a^\dagger_{k}b^\dagger_{-k}\right)\left|0\right>
\end{equation}
Commuting $\exp\left(-\sum\limits_k \kappa(k) (a^\dagger_k a_k + b^\dagger_k b_k)\right)$ \, through \, $\exp\left(\sum\limits_k \sgn(k) a^\dagger_kb^\dagger_{-k}\right)$, we get 
$$
\exp\left(\sum\limits_k \sgn(k) e^{-2\kappa(k)} a^\dagger_kb^\dagger_{-k}\right)\left|0\right>$$ 

Thus,
\begin{eqnarray}
&& \text{sgn}(k)\gamma(k) = e^{-2\kappa(k)} \sgn(k) \nonumber\\
&\Rightarrow & \kappa(k) = -\frac{1}{2}\log(\gamma(k))\
\label{forbch}
\end{eqnarray}
\section{\label{refW4}Refermionization of bosonic bilinear $\mathcal{W}_4$}
The bosonic(real scalar) bilinear $\mathcal{W}_4$ current \cite{Bakas:1990ry, Pope:1991ig} is
\begin{eqnarray}
 \mathcal{W}_4(w)=2\partial\phi\partial^3\phi-3\partial^2\phi\partial^2\phi\
\end{eqnarray}
Using U(1) current relation $J=\psi^\dagger\psi=\frac{i}{\sqrt{4\pi}}\,\partial\phi$ and normal ordering gives the refermionized $\mathcal{W}_4$ current. Because of the fermionic anti-commutation relation most of the four fermion terms drop out and the only four fermion term that survives is $\partial\psi^{\dagger}\partial\psi\psi^{\dagger}\psi$. Finally, the expression is
\begin{eqnarray}
 \mathcal{\tilde{W}}_4(w)=\frac{7i}{6}\,\psi^{\dagger}\partial^3\psi+\frac{3i}{2}\,\partial^2\psi^{\dagger}\partial\psi-\frac{3i}{2}\,\partial\psi^{\dagger}\partial^2\psi-\frac{7i}{6}\,\partial^3\psi^{\dagger}\psi-2\,\partial\psi^{\dagger}\partial\psi\psi^{\dagger}\psi\
 \label{newW4}
\end{eqnarray}
And the corresponding charge is
\begin{align}
&\tilde{W}_4=\frac{1}{4\pi}\left(\frac{14}{3}\int_{-\infty}^{\infty}\frac{dk}{2\pi}\,|k|^3\left[a^{\dagger}_k a_k+b^{\dagger}_k b_k\right]\right.\nonumber\\
 &\quad+2\int_{-\infty}^{\infty}\frac{dk_1dk_2dk_3dk_4}{(2\pi)^4}\,|k_1||k_2|\left[a^{\dagger}_{k_1}a_{k_2}a^{\dagger}_{k_3}a_{k_4}\delta(k_1-k_2+k_3-k_4)+a^{\dagger}_{k_1}a_{k_2}b_{k_3}b^{\dagger}_{k_4}\delta(k_1-k_2-k_3+k_4)\right.\nonumber\\
 &\qquad\qquad-a^{\dagger}_{k_1}b^{\dagger}_{k_2}b_{k_3}a_{k_4}\delta(k_1+k_2-k_3-k_4)-b_{k_1}a_{k_2}a^{\dagger}_{k_3}b^{\dagger}_{k_4}\delta(-k_1-k_2+k_3+k_4)\nonumber\\
 &\qquad\qquad+\left.b_{k_1}b^{\dagger}_{k_2}a^{\dagger}_{k_3}a_{k_4}\delta(-k_1+k_2+k_3-k_4)+b_{k_1}b^{\dagger}_{k_2}b_{k_3}b^{\dagger}_{k_4}\delta(-k_1+k_2-k_3+k_4)\right]\Bigg{)}\
 \label{newW4ch}
\end{align}

\bibliography{thermal} 
\bibliographystyle{JHEP}

\end{document}